\documentclass[twocolumn,english,aps,pre,showpacs]{revtex4}
\usepackage{amsmath}
\usepackage{amssymb}
\usepackage{graphicx}

\makeatletter
\usepackage{babel}
\makeatother

\begin{document}

\title{Waves at surfactant-laden liquid--liquid crystal interface}

\author{S.~V.~Lishchuk}

\affiliation{
Materials and Engineering Research Institute, Sheffield Hallam University,\\
Howard Street, Sheffield S1~1WB, United Kingdom
}


\begin{abstract}
A theoretical study is presented of surface waves at a monomolecular surfactant
film between an isotropic liquid and a nematic liquid crystal for the case when
the surfactant film is in the isotropic two-dimensional fluid phase and induces
homeotropic (normal to the interface) orientation of the nematic director. The
dispersion relation for the surface waves is obtained, and different surface
modes are analyzed with account being taken of the anchoring induced by the
surfactant layer, the curvature energy of the interface, and the anisotropy of
the viscoelastic coefficients.  The dispersion laws for capillary and
dilatational surface modes retain structure similar to that in isotropic
systems, but involve anisotropic viscosity coefficients. Additional modes are
related to relaxation of the nematic director field due to anchoring at the
interface. The results can be used to determine different properties of
nematic-surfactant-isotropic interfaces from experimental data on surface light
scattering.
\end{abstract}

\maketitle


\section{\label{sec:Introduction}Introduction}

The presence of a surfactant film at a fluid-fluid interface alters the
dynamics of the interface. This is manifested in behavior of the interfacial
waves, induced either externally or by thermal fluctuations
\cite{Schriven:1960-98,Kats:1988-940,Buzza:2002-8418}. The interfacial dynamics
can be probed by measuring the light scattered on such surface waves (see the
review by Earnshaw \cite{Earnshaw:1996-1}). The scattering of light on surface
waves is a powerful tool for probing the properties of surfactant films at
fluid interfaces \cite{Kramer:1971-2097,Earnshaw:1995-1087,Buzza:2002-8418}, and
a variety of systems have been recently investigated using this method
(e.g.\ refs~\cite{Milling:2001-5305,Yang:2001-6254,Eastoe:2003-7734,
Wang:2004-2930,Sakai:2005-063908,Rojas:2005-22440,Kim:2006-4889}, see also the
review by Cicuta and Hopkinson \cite{Cicuta:2004-97}).

Recently, the application of surfactant films to modify the interfacial
properties has been extended to the systems in which one of the fluids is in
liquid-crystalline phase (e.g.\ liquid crystal colloids
\cite{Poulin:1997-1770}). The presence of a liquid crystal as one of the fluids
complicates the problem of probing the interfacial properties by studying the
dynamics of the surface waves for the following reasons. Firstly, there are
additional degrees of freedom in the bulk of the liquid crystal phase due to its
anisotropy. Secondly, the interaction with the surfactant film is more
complicated due to anisotropic anchoring. Finally, the surfactant film in the
anisotropic field created by the neighboring liquid crystal can itself show
anisotropic behavior, even if it behaves as an two-dimensional isotropic fluid
at the boundary between isotropic fluids.

A promising new direction for chemical and biological sensing devices has
recently emerged which utilizes the properties of surfactant films
self-assembled on the interface between water and a nematic liquid crystal. The
surfactant film induces preferred orientation of the nematic director
\cite{Crawford:1996-3647,Fazio:2001-061712,Bahr:2006-030702}. The adsorption of
chemical or biological molecules at such interface can then lead to
reorientation of the nematic director, enabling detection by an imaging system
\cite{Brake:2002-6101,Brake:2003-6436, Clare:2005-6451, Lockwood:2005-111,
Lockwood:2005-6805}.

In these methods, easy detection is limited to the systems in which adsorption
changes anchoring properties of the interface with respect to the adjacent
liquid crystal phase quite considerably. Namely, the equilibrium anchoring angle
should change in magnitude. The range of application of these systems could be
made significantly broader, however, if a method were used that was sensitive to
changes in the anchoring properties of the interface that did not necessarily
result in nematic director reorientation. For example, the anchoring orientation
may remain unchanged \cite{Poulin:1997-1770,Brake:2002-6101}, the adsorption
only changing the strength of the anchoring.

If a small amount of an analyte is present in the water it may be adsorbed at
the surfactant layer, provided the surfactant molecules possess appropriate
chemical properties. Generally, such adsorption will result in a change in the
elastic and viscous properties of the interface. Hence sensitive experiments
which are able to determine the interfacial properties will allow much more
detailed experimental insight into the properties of the interaction between the
surfactants and the analyte than has hitherto been available, and experimental
study of surface waves is a possible technique for this purpose.

The theoretical description of surface waves at interfaces between nematic and
isotropic liquids was made back in 1970s
\cite{Langevin:1972-101,Hayes:1975-115,Parsons:1974-2341}. The results
demonstrated that the spectrum of surface waves has a more complicated structure
than in the isotropic case, and allows the use surface scattering experiments to
determine properties of nematic interfaces
\cite{Langevin:1972-249,Langevin:1973-317,McQueen:1974-1983,Langevin:1975-745,
Sohl:1980-1256}. Since then, several theoretical and experimental advances have
been made, and presently these systems remain a subject of investigation
\cite{PopaNita:2002-041703,PopaNita:2003-061707, PopaNita:2005-061706,
Elgeti:2005-407,Wolfsheimer:061703}.

The present paper presents a theoretical study of the dispersion of the
surface waves at a monomolecular surfactant film between an isotropic liquid
(e.g.\ water) and a nematic liquid crystal.The main distinguishing features of
such interfaces, are
(i) the anchoring induced by the surfactant layer,
(ii) the curvature energy of the interface,
(iii) reduction of surface tension due to surfactant, and
(iv) the anisotropy of the surface viscoelastic coefficients.
We base our treatment on the mechanical model for anisotropic curved interfaces
by Rey \cite{Rey:2006-219}, which takes into account anchoring and bending
properties of the surfactant. We consider the case of the insoluble surfactant
film that is in its most symmetric phase (isotropic two-dimensional fluid), and
induces homeotropic (normal to the surface) orientation of the director.

The paper is organized as follows. The continuum model used in the rest of the
paper is set up in Section~\ref{sec:Model}. In Section~\ref{sec:Dispersion} the
dispersion relation for surface waves is derived. In Section~\ref{sec:Modes} the
numerical solution of the dispersion relation is solved with typical values of
material parameters, and dispersion laws for different surface modes are
analyzed in absence of the external magnetic field, and the influence of the
magnetic field is discussed in Section~\ref{sec:Field}.

The explicit form of the dispersion relation is written in
Appendix~\ref{app:Dispersion}.


\section{\label{sec:Model}The model}

In this section we formulate the model of the surfactant-laden interface between
an isotropic liquid and a nematic liquid crystal, used in the present paper, and
write down the governing equations. We base our treatment upon the models of the
nematic-isotropic interface by Rey \cite{Rey:2006-219,Rey:2000-1540}, and well
known hydrodynamic description of isotropic liquids \cite{Landau:FM} and nematic
liquid crystals \cite{Gennes:PLC,Landau:TE}.

We consider the case when the surfactant film induces homeotropic (normal to the
surface) orientation of the nematic director, which is usually true in a range
of the surfactant concentrations
\cite{Poulin:1997-1770,Gupta:1998-2077,Brake:2002-6101}.
This case is the simplest to analyze, and, at the same time, the most important
for biosensing applications where the direct change in anchoring angle cannot be
always observed.

We include optional external magnetic field in our study and limit our analysis
by considering the direction of the magnetic field that does not change
equilibrium orientation of the nematic director.

We assume that the system is far enough from any phase transitions both in the
surfactant film \cite{Kaganer:1999-779} and in the nematic phase
\cite{Gennes:PLC}. Thus we avoid complications related to the fluctuations
of the nematic and surfactant order parameters and the divergence of
viscoelastic parameters near phase transitions.

The surfactant films can exhibit rich phase behavior \cite{Kaganer:1999-779},
and the form of the surface stress tensor depends upon the symmetry of the
interface. However, this does not normally influence much the dispersion laws of
the surface modes compared to the isotropic case \cite{Kats:1988-940}. In the
present paper we assume that the surfactant film is in the most symmetric phase
(isotropic two-dimensional fluid). Although the symmetry of the film should
break in presence of the adjacent liquid-crystalline bulk phase, the film
remains isotropic in equilibrium if the anchoring of the nematic is homeotropic,
and symmetry breaking can occur only due to fluctuations of the director field.
If we introduce the order parameter for the film, the corresponding anisotropic
contributions to the interfacial stress tensor would be of higher order in the
fluctuations of the dynamic variables than is required in our linearized
treatment, so such contributions can be omitted.

We consider a surfactant layer at an interface between nematic and isotropic
liquids to be macroscopically infinitely thin. We assume that the surfactant
film is insoluble and Newtonian. This means that the model is applicable to
systems in which the interchange of surfactant molecules between the interface
and adjacent bulk fluids is small, and the relaxation of the orientation of
surfactant molecules is fast compared to relaxation of surface waves. We also
assume heat diffusion to be sufficiently fast so that the system is in thermal
equilibrium. We do not consider systems where other effects, such as polarity,
are important.

We shall choose coordinate system in such a way that the unperturbed interface
lies at a plane $z=0$, the half-space $z<0$ is occupied by the uniaxial nematic
liquid crystal, and the half-space $z>0$ is filled by the isotropic liquid.
Other details of the geometry used in the present paper are summarized in
Appendix~\ref{app:Geometry}.

The central equations in the present Section are the conditions for the balance
of forces (Eq.~(\ref{eq:forceBalance}) and torques (Eq.~(\ref{eq:torqueBalance})
at the interface. The explicit form of these equations depends upon the chosen
macroscopic model, and the rest of this Section is devoted to formulation of the
model used in the present paper.


\subsection{Balance equations}

The interfacial force balance equation is the balance between the interfacial
force and the bulk stress jump:
\begin{equation}
\label{eq:forceBalance}
\mathbf F^S+\mathbf F^N+\mathbf F^I=0.
\end{equation}
Here
\begin{equation}
\label{eq:def-ForceS}
\mathbf F^S=\nabla_s\cdot\boldsymbol\Sigma^S
\end{equation}
is the force per unit area exerted by the interfacial stress
$\boldsymbol\Sigma^S$,
\begin{equation}
\label{eq:def-ForceI}
\mathbf F^I=\left.\boldsymbol\Sigma^I\right|_s\cdot\mathbf k
\end{equation}
is the force per unit area exerted by the isotropic fluid,
\begin{equation}
\label{eq:def-ForceN}
\mathbf F^N=-\left.\boldsymbol\Sigma^N\right|_s\cdot\mathbf k
\end{equation}
is the force per unit area exerted by the nematic liquid crystal, the subscript
$s$ indicates that the bulk stress fields in the isotropic liquid,
$\boldsymbol\Sigma^I$, and in the nematic, $\boldsymbol\Sigma^N$, are evaluated
at the interface, $\mathbf k$ is the unit vector normal to the interface and
directed into the isotropic liquid.

The interfacial torque balance equation can be cast as
\begin{equation}
\label{eq:torqueBalance}
\mathbf T^S+\mathbf T^N=0,
\end{equation}
where $\mathbf T^S$ is the interfacial torque arising due to surface
interactions, $\mathbf T^N$ is the torque exerted upon the interface by the
adjacent nematic liquid crystal.

The explicit model for surface and bulk stresses and torques that enter
Eqs~(\ref{eq:forceBalance}) and (Eq.~(\ref{eq:torqueBalance}) is expanded in
the remainder of this section.


\subsection{Surface elastic stress and torque}

In this and the following subsections we summarize the equations for the surface
stress tensor $\boldsymbol\Sigma^S$ and surface torque vector $\mathbf T^S$. We
represent these quantities as a sum of corresponding non-dissipative (elastic)
and dissipative (viscous) contributions:
\begin{equation}
\label{eq:sigmaS}
\boldsymbol\Sigma^S=\boldsymbol\Sigma^{Se}+\boldsymbol\Sigma^{Sv},
\end{equation}
\begin{equation}
\mathbf T^S=\mathbf T^{Se}+\mathbf T^{Sv}.
\end{equation}
To describe the non-dissipative contributions in the surface stress tensor,
$\boldsymbol\Sigma^{Se}$, and surface torque vector, $\mathbf T^{Se}$, we use
the equilibrium model proposed by Rey \cite{Rey:2006-219}, which is summarized
below.

Rey considered the interface with the Helmholtz free energy per unit mass
$\mathcal F^S$ of the form
\begin{equation}
\mathcal F^S=\mathcal F^S\left(\rho^S,\mathbf k,\mathbf b\right),
\end{equation}
where $\rho^S$ is the surface mass density, $\mathbf b$ is the second
fundamental tensor of the interface (see Appendix~\ref{app:Geometry}). The
corresponding differential was written as
\begin{equation}
d\mathcal F^S=-\frac\sigma{\left(\rho^S\right)^2}d\rho^S
+\frac{\boldsymbol\xi_\Vert}\rho\cdot d\mathbf k
+\frac{\mathbf M}\rho:d\mathbf b,
\end{equation}
where
\begin{equation}
\label{eq:def-tau}
\sigma=\left[
-\left(\rho^S\right)^2\frac{\partial\mathcal F^S}{\partial\rho^S}
\right]_{\mathbf k,\mathbf b}
\end{equation}
is the interfacial tension,
\begin{equation}
\label{eq:def-xi}
\boldsymbol\xi_\Vert=
\left(\rho^S\mathbf I_S\cdot\frac{\partial\mathcal F^S}{\partial\mathbf k}\right)
_{\rho^S,\mathbf b}
\end{equation}
is the tangential component of the capillary vector ($\mathbf I_S$ is the
surface projector), and
\begin{equation}
\label{eq:def-M}
\mathbf M=\left(\rho^S\frac{\partial\mathcal F^S}{\partial\mathbf b}\right)
_{\rho^S,\mathbf k}
\end{equation}
is the bending moment tensor. The elastic surface stress tensor was found to
be
\begin{equation}
\label{eq:sigmaSe}
\boldsymbol\Sigma^{Se}=
\sigma\mathbf I_S-\mathbf M\cdot\mathbf b+\mathbf h^{Se}_\Vert\mathbf k,
\end{equation}
where the tangential surface molecular field is given by
\begin{equation}
\label{eq:def-hSe}
\mathbf h^{Se}_\Vert=-\mathbf I_S\cdot\frac{\delta\mathcal F^S}{\delta\mathbf k}
=-\boldsymbol\xi_\Vert-\mathbf I_S\cdot\left(\nabla_s\cdot\mathbf M\right),
\end{equation}
$\nabla_s$ is surface gradient operator, $\delta/\delta\mathbf k$ denotes
variational derivative with respect to $\mathbf k$. The elastic contribution
to surface torque was written as
\begin{equation}
\mathbf T^{Se}=
-\boldsymbol\epsilon:\boldsymbol\Sigma^{Se}+\nabla\cdot\mathbf C_S,
\end{equation}
where
\begin{equation}
\mathbf C_S=-\mathbf M\cdot\boldsymbol\epsilon_S
\end{equation}
is the surface couple stress, $\boldsymbol\epsilon$ is the Levi-Civita tensor,
and $\boldsymbol\epsilon_S=-\mathbf I_S\times\mathbf k$ is the surface
alternator tensor.


\subsection{Surface viscous stress and torque}

The viscous properties of interfaces between an isotropic fluid and a nematic
liquid crystal were considered in detail by Rey \cite{Rey:2000-1540}, and the
results are summarized below.

The forces and fluxes that contribute to the dissipation function $R$ were
identified as follows:
\begin{eqnarray}
\nonumber
R&=&\boldsymbol\Sigma^{Sv}_s:\mathbf S^S
+\boldsymbol\Sigma^{Sv}_a:\mathbf A^S
\\
\label{eq:EntropyProduction}
&+&\mathbf h^{Sv}_\Vert\cdot\left(\mathbf I_s\cdot\frac{d\mathbf n_\Vert}{dt}\right)
+\mathbf h^{Sv}_\bot\cdot\left(\mathbf{kk}\cdot\frac{d\mathbf n_\bot}{dt}\right),
\end{eqnarray}
where $\boldsymbol\Sigma^{Sv}_s$ and $\boldsymbol\Sigma^{Sv}_a$ are,
correspondingly, symmetric and antisymmetric parts of the surface viscous stress
tensor $\boldsymbol\Sigma^{Sv}$, $\mathbf h^{Sv}_\Vert$ and
$\mathbf h^{Sv}_\bot$ are the components of the surface viscous molecular field
tangential and normal to the surface,
\begin{equation}
\label{eq:SS}
\mathbf S^S=\frac12\left[\nabla_s\mathbf v^S\cdot\mathbf I_s
+\mathbf I_s\cdot\left(\nabla_s\mathbf v^S\right)^\textrm T\right]
\end{equation}
is the surface rate-of-deformation tensor ($(\cdots)^\textrm T$ denotes the
transposed tensor),
\begin{equation}
\mathbf A^S=\frac12\left[\nabla_s\mathbf v^S\cdot\mathbf I_s
-\mathbf I_s\cdot\left(\nabla_s\mathbf v^S\right)^\textrm T\right]
\end{equation}
is the surface vorticity tensor, $\mathbf v^S$ is surface velocity,
\begin{equation}
\frac{d\mathbf n_\Vert}{dt}=\frac{\partial\mathbf n_\Vert}{\partial t}
+\mathbf v^S\cdot\nabla_s\mathbf n_\Vert
\end{equation}
and
\begin{equation}
\frac{d\mathbf n_\bot}{dt}=\frac{\partial\mathbf n_\bot}{\partial t}
+\mathbf v^S\cdot\nabla_s\mathbf n_\bot
\end{equation}
are the total time derivatives of the components
$\mathbf n_\Vert=\mathbf I_s\cdot\mathbf n$ and
$\mathbf n_\bot=\mathbf{kk}\cdot\mathbf n$ of the nematic director field
$\mathbf n$, tangential and normal to the surface, correspondingly.

Generally, presence of the surfactant film at the interface complicates the
form of the entropy production due to additional internal degrees of freedom of
the surfactant, and to the anisotropy of the adjacent nematic liquid. However,
if the surfactant film that is in its isotropic liquid phase and favors
homeotropic anchoring of the nematic, the resulting anisitropic terms in the
entropy production introduce corrections to the hydrodynamic equations of higher
order than linear, and therefore can be neglected in the linearized treatment.
Since this is the case we are considering, we shall adopt the form of the
entropy production (\ref{eq:EntropyProduction}) in our model and use the form of
the viscous contribution to the surface stress tensor derived by Rey
\cite{Rey:2000-1540}, which is given by
\begin{eqnarray}
\nonumber
\boldsymbol\Sigma^{Sv}&=&\alpha^S_1
\mathbf S^S:\mathbf n_\Vert\mathbf n_\Vert\mathbf n_\Vert\mathbf n_\Vert
\\
\nonumber
&+&\alpha^S_2\mathbf n_\Vert\mathbf N^S
+\alpha^S_3\mathbf N^S\mathbf n_\Vert
+\alpha^S_4\mathbf S^S
\\
\nonumber
&+&\alpha^S_5\mathbf n_\Vert\mathbf n_\Vert\cdot\mathbf S^S
+\alpha^S_6\mathbf S^S\cdot\mathbf n_\Vert\mathbf n_\Vert
\\
\nonumber
&+&\alpha^S_7\mathbf n_\Vert\mathbf n_\Vert
\left(\mathbf n_\Vert\cdot\mathbf N^S\right)
+\beta^S_1\mathbf I_s\left(\mathbf I_s:\mathbf S^S\right)
\\
&+&\beta^S_2\left[
\mathbf n_\Vert\mathbf n_\Vert\left(\mathbf I_s:\mathbf S^S\right)
+\mathbf I_s\left(\mathbf n_\Vert\mathbf n_\Vert:\mathbf S^S\right)
\right],
\end{eqnarray}
where $\mathbf N^S$ is the surface Jaumann (corrotational) derivative
\cite{Thiffeault:2001-5875} of the tangential component of the director
$\mathbf n_\Vert$, and $\alpha^S_{1{-}7}$, $\beta^S_{1{-}2}$ are nine
independent surface viscosity coefficients. In the isotropic case $\mathbf n=0$,
the expression for the surface viscous stress tensor reduces to the viscous
stress tensor of Boussinesq-Schriven surface fluid
\cite{Schriven:1960-98,Aris:VT} with the interfacial shear viscosity $\eta_s$
given by
\begin{equation}
\eta_s=\frac{\alpha^S_4}2,
\end{equation}
and dilatational viscosity $z\eta_s$ given by
\begin{equation}
\zeta_s=\frac{\alpha^S_4}2+\beta^S_1.
\end{equation}

The surface viscous torque, corresponding to Eq.~(\ref{eq:EntropyProduction}),
is given by \cite{Rey:2000-1540}
\begin{equation}
\mathbf T^{Sv}=-\mathbf n\times\mathbf h^{Sv},
\end{equation}
where the surface viscous molecular field $\mathbf h^{Sv}$ is
\begin{eqnarray}
\nonumber
\mathbf h^{Sv}&=&\gamma^S_2\mathbf A\cdot\mathbf n_\Vert
+\gamma^S_{1\Vert}\mathbf N^S
+\alpha^S_6\mathbf n_\Vert
\left(\mathbf n_\Vert\mathbf n_\Vert:\mathbf A^S\right)
\\
\label{eq:def-hSv}
&+&\frac{\gamma^S_2}2\mathbf n_\Vert\left(\mathbf I_s:\mathbf S^S\right)
+\gamma^S_{1\bot}\mathbf k\mathbf k\cdot\frac{d\mathbf n_\bot}{dt}.
\end{eqnarray}
The viscosity coefficients $\gamma^S_i$ can be expressed in terms of quantities
$\alpha^S_i$. We shall need only the expression for the tangential rotational
viscosity:
\begin{equation}
\gamma^S_{1\Vert}=\alpha^S_3-\alpha^S_2.
\end{equation}


\subsection{Anchoring and curvature energies}

To calculate explicitly the interfacial tension $\sigma$
(Eq.~(\ref{eq:def-tau})), the tangential component of the capillary vector
$\boldsymbol\xi_\Vert$ (Eq.~(\ref{eq:def-xi})), and the bending moment tensor
$\mathbf M$ (Eq.~(\ref{eq:def-M})), we need
to know the dependence of the surface free energy $\mathcal F^S$ on the 
orientation of the interface given by unit normal vector $\mathbf k$, and on
its curvature described by second fundamental tensor $\mathbf b$. For small
deviations of $\mathbf k$ and $\mathbf b$ from equilibrium, we can expand the
free energy in powers of these quantities and truncate the series. The result
can be represented as
\begin{equation}
\label{eq:FS}
\mathcal F^S(\rho^S,\mathbf k,\mathbf b)
=\mathcal F^S_t(\rho^S)
+\mathcal F^S_a(\mathbf k)
+\mathcal F^S_c(\mathbf b),
\end{equation}
each of the contribution described below.

The contribution $\mathcal F^S_t$ corresponds to the surface tension
$\bar\sigma$ of the equilibrium interface (flat interface, adjacent nematic
director normal to the interface):
\begin{equation}
\label{eq:FSt}
\rho^S\mathcal F^S_t=\bar\sigma.
\end{equation}

The anchoring contribution to the surface free energy density, $\mathcal F^S_a$,
describes the energetics of the preferred alignment direction of the nematic
director relative to the interface. For the homeotropic equilibrium anchoring,
it can be written in terms of $\mathbf n_\Vert$ as follows:
\begin{equation}
\label{eq:FSa}
\rho^S\mathcal F^S_a=\frac12W\mathbf n_\Vert^2+o\left(\mathbf n_\Vert^2\right).
\end{equation}
Such expansion applied to the widely used Rapini-Papoular form of the anchoring
free energy density \cite{Rapini:1969-54}
\begin{equation}
\rho^S\mathcal F_{RP}=\frac{W_{RP}}2\left(\mathbf n\cdot\mathbf k\right)^2,
\end{equation}
shows that these definitions of the anchoring strength coefficient have opposite
signs:
\begin{equation}
W=-W_{RP}.
\end{equation}
We shall use $W$ as the anchoring strength coefficient to ensure that it is
positive in the case of the homeotropic anchoring being considered.

The third contribution to the surface free energy density, $\mathcal F^S_c$, is
caused by finite interface thickness, and is related to the difference of the
curvature of a surfactant film from the locally preferred (spontaneous) value.
The widely used form of this contribution is the Helfrich curvature expansion
\cite{Helfrich:1973-693,Safran:STSIM}
\begin{equation}
\label{eq:FSc}
\rho^S\mathcal F^S_c=-2\kappa\bar H^2+2\kappa\left(H-\bar H\right)^2+\bar\kappa K
+o\left((\nabla\mathbf k)^2\right).
\end{equation}
Here the geometry of the interface is described by the mean curvature $H$ and
the Gaussian curvature $K$, and the material parameters characterizing the
interface are the bending rigidity $\kappa$, the saddle-splay (or Gaussian)
rigidity $\bar\kappa$, and the spontaneous curvature $\bar H$. The term
$-2\kappa\bar H^2$ guarantees that the curvature energy of a flat interface
($H=0$, $K=0$) is zero.


\subsection{Surfactant concentration}

To complete the description of the interface, we need the continuity equation
for the surfactant concentration $\nu$.
For insoluble surfactants, the continuity equation reads:
\begin{equation}
\label{eq:continuityS}
\frac{d\nu}{dt}+\nu\nabla_s\cdot\mathbf v^S=0.
\end{equation}

We shall extend the description of the dependence of the interfacial tension
upon the concentration of surfactant, presented by Buzza \cite{Buzza:2002-8418},
to other parameters characterizing the interface (surface tension $\bar\sigma$,
anchoring strength $W$, bending rigidity $\kappa$, saddle-splay rigidity
$\bar\kappa$, spontaneous curvature $\bar H$, and surface viscosities $\alpha^S_i$,
$\beta^S_i$). For small deviation $\delta\nu=\nu-\nu_0$ of the surfactant
concentration $\nu$ from its equilibrium value $\nu_0$, these coefficients can
be written in form
\begin{equation}
\bar\sigma(\nu)=
\bar\sigma(\nu_0)+\frac{\partial\bar\sigma}{\partial\nu}\delta\nu,
\end{equation}
and similarly for other quantities. Casting surface velocity $\mathbf v^S$ as
the time derivative of the small surface displacement $\mathbf u$,
\begin{equation}
\mathbf v^S=\frac{d\mathbf u}{dt},
\end{equation}
we obtain from the continuity equation Eq.~(\ref{eq:continuityS}) that
\begin{equation}
\delta\nu=-\nu_0\nabla_s\cdot\mathbf u.
\end{equation}
This allows us to represent the material parameters of the interface as
\begin{equation}
\bar\sigma(\nu)=\sigma_0+\epsilon_0\nabla_s\cdot\mathbf u,
\end{equation}
\begin{equation}
W(\nu)=W_0+W_1\nabla_s\cdot\mathbf u,
\end{equation}
\begin{equation}
\kappa(\nu)=\kappa_0+\kappa_1\nabla_s\cdot\mathbf u,
\end{equation}
\begin{equation}
\bar H(\nu)=\bar H_0+\bar H_1\nabla_s\cdot\mathbf u.
\end{equation}
In these formulas $\sigma_0=\bar\sigma(\nu_0)$, $W_0=W(\nu_0)$,
$\kappa_0=\kappa(\nu_0)$, $\bar H_0=\bar H(\nu_0)$ are, correspondingly, the
interfacial tension, anchoring strength, bending rigidity, and spontaneous
curvature in the unperturbed interface,
$\epsilon_0=-\nu_0\partial\bar\sigma/\partial\nu$ is the static dilatational
elasticity, $W_1=-\nu_0\partial W/\partial\nu$, $\kappa_1=-\nu_0\partial\kappa/\partial\nu$, and
$\bar H_1=-\nu_0\partial\bar H/\partial\nu$ are coefficients in the first order
term of the expansion of anchoring strength, bending rigidity, and spontaneous
curvature in powers of ($\nabla_s\cdot\mathbf u$). There are similar expansions
for Gaussian rigidity $\bar\kappa$ and surface viscosities
$\alpha^S_i$, $\beta^S_i$.


\subsection{Magnetic field}

Magnetic field $\boldsymbol{\mathcal H}$ in the isotropic and nematic regions
satisfies Maxwell equations \cite{Landau:ECM,Hayes:1975-115}
\begin{equation}
\label{eq:rotMaxwell}
\nabla\times\boldsymbol{\mathcal H}=0,
\end{equation}
\begin{equation}
\label{eq:divMaxwell}
\nabla\cdot\boldsymbol{\mathcal H}=0.
\end{equation}
Neglecting magnetization of the interface, the boundary conditions read
\begin{equation}
\left.\mathbf I_s\cdot\boldsymbol{\mathcal H}\right|_N=
\left.\mathbf I_s\cdot\boldsymbol{\mathcal H}\right|_I,
\end{equation}
\begin{equation}
\left.\mathbf k\cdot(\boldsymbol{\mathcal H}+4\pi\boldsymbol{\mathcal M})\right|_N=
\left.\mathbf k\cdot(\boldsymbol{\mathcal H}+4\pi\boldsymbol{\mathcal M})\right|_I.
\end{equation}
Here the magnetization of the isotropic liquid is
\begin{equation}
\left.\boldsymbol{\mathcal M}\right|_I=\left.\chi^I\boldsymbol{\mathcal H}\right|_I,
\end{equation}
where $\chi^I$ is the magnetic permeability of the isotropic liquid, the
magnetization of the uniaxial nematic liquid crystal is \cite{Gennes:PLC}
\begin{equation}
\left.\boldsymbol{\mathcal M}\right|_N=
\left.\chi^N_\bot\boldsymbol{\mathcal H}
+\chi_a(\boldsymbol{\mathcal H}\cdot\mathbf n)\mathbf n\right|_N,
\end{equation}
where $\chi_a$ is the difference of the longitudinal and transversal magnetic
permeabilities of the nematic:
\begin{equation}
\chi_a=\chi^N_\Vert-\chi^N_\bot.
\end{equation}


\subsection{Isotropic liquid}

We assume both the isotropic liquid and the nematic liquid crystal are
incompressible, so that their densities $\rho^I$ and $\rho^N$, are constant.

The linearized equations for the incompressible isotropic liquid are well known
\cite{Landau:FM}. They are the continuity equation
\begin{equation}
\label{eq:continuity}
\nabla\cdot\mathbf v=0,
\end{equation}
and Navier-Stokes equations
\begin{equation}
\label{eq:Navier-Stokes}
\rho^I\frac{\partial\mathbf v}{\partial t}=\nabla\cdot\boldsymbol\Sigma^I,
\end{equation}
where the hydrodynamic stress tensor is given by
\begin{equation}
\label{eq:SigmaI}
\boldsymbol\Sigma^I=-p\mathbf I+2\eta\mathbf S,
\end{equation}
where $\eta$ is the shear viscosity of the isotropic liquid, $\mathbf I$ is
the unit tensor,
\begin{equation}
\label{eq:S}
\mathbf S=
\frac12\left[\nabla\mathbf v+\left(\nabla\mathbf v\right)^\textrm{T}\right]
\end{equation}
is the strain rate tensor.

We assume the non-slip boundary condition for the velocities of bulk fluids
adjacent to the interface, which means the equality of the velocity of
surfactant, $\mathbf v^S$, and that of the bulk fluids at an interface,
$\left.\mathbf v\right|_s$:
\begin{equation}
\mathbf v^S=\left.\mathbf v\right|_s.
\end{equation}


\subsection{Nematic liquid crystal}

To describe the dynamics of the nematic liquid crystal that is far from the
isotropic-nematic transition and has small deviations from its equilibrium
state, we shall use the linearized form of the Eriksen-Leslie theory
\cite{Forster:1971-1016,Landau:TE,Gennes:PLC}. The linearized equations for the
incompressible nematic liquid crystal are the continuity equation
(\ref{eq:continuity}), the equation for the velocity
\begin{equation}
\label{eq:EriksenLeslie}
\rho^N\frac{\partial\mathbf v}{\partial t}=\nabla\cdot\boldsymbol\Sigma^N,
\end{equation}
and the equation for the director
\begin{equation}
\label{eq:dndt}
\frac{\partial\delta\mathbf n}{\partial t}=\mathbf n_0\cdot\mathbf A
+\lambda\left(\mathbf I-\mathbf n_0\mathbf n_0\right)
\cdot\mathbf S\cdot\mathbf n_0+\frac1{\gamma_1}\mathbf h.
\end{equation}
Here
\begin{equation}
\mathbf A=
\frac12\left[\nabla\mathbf v-\left(\nabla\mathbf v\right)^\textrm{T}\right]
\end{equation}
is the antisymmetric vorticity tensor, $\lambda$ is the reactive material
parameter, $\gamma_1$ is orientational viscosity, $\mathbf h$ is the molecular
field which, assuming Frank form of the elastic free energy of a nematic liquid
crystal in magnetic field \cite{Gennes:PLC}
\begin{eqnarray}
\nonumber
\mathcal F_F&=&\frac{K_1}2\left(\nabla\cdot\mathbf n\right)^2
+\frac{K_2}2\left[\mathbf n\cdot\left(\nabla\times\mathbf n\right)\right]^2
\\
&+&\frac{K_3}2\left[\mathbf n\times\left(\nabla\times\mathbf n\right)\right]^2
-\frac12\chi_a(\mathbf n\cdot\boldsymbol{\mathcal H})^2,
\end{eqnarray}
has the linearized form
\begin{equation}
\label{eq:h}
\mathbf h=\left(\mathbf I-\mathbf n_0\mathbf n_0\right)\cdot\mathbf h^\ast,
\end{equation}
where
\begin{eqnarray}
\label{eq:h-star}
\mathbf h^\ast=K_1\nabla\nabla\cdot\delta\mathbf n
-K_2\nabla\times\left\{\mathbf n_0\left[\mathbf n_0\cdot
\left(\nabla\times\delta\mathbf n\right)\right]\right\}+
\\
+K_3\nabla\times\left\{\mathbf n_0\times\left[\mathbf n_0\times
\left(\nabla\times\delta\mathbf n\right)\right]\right\}
+\chi_a(\mathbf n\cdot\boldsymbol{\mathcal H})\boldsymbol{\mathcal H},
\end{eqnarray}
$K_1$, $K_2$, and $K_3$ are the splay, twist, and bend Frank elastic constants,
correspondingly.

The stress tensor can be represented as a sum of reactive and viscous
(dissipative) contributions,
\begin{equation}
\label{eq:SigmaN}
\boldsymbol\Sigma^N=\boldsymbol\Sigma^{Nr}+\boldsymbol\Sigma^{Nv}.
\end{equation}
The linearized form of the reactive part is
\begin{equation}
\label{eq:SigmaNr}
\boldsymbol\Sigma^{Nr}=-p\mathbf I
+\frac12\left(\mathbf n_0\mathbf h-\mathbf h\mathbf n_0\right)
-\frac\lambda2\left(\mathbf n_0\mathbf h+\mathbf h\mathbf n_0\right).
\end{equation}
The linearized viscous stress tensor of incompressible nematic is
\begin{eqnarray}
\nonumber
\boldsymbol\Sigma^{Nv}&=&2\nu_2\mathbf S+2\left(\nu_3-\nu_2\right)
\left(\mathbf n_0\mathbf S\cdot\mathbf n_0
+\mathbf n_0\cdot\mathbf S\mathbf n_0\right)
\\
\label{eq:SigmaNv}
&+&2\left(\nu_1+\nu_2-2\nu_3\right)
\mathbf n_0\mathbf n_0\mathbf n_0\mathbf n_0:\mathbf S.
\end{eqnarray}
The quantities $\nu_1$, $\nu_2$, $\nu_3$, $\gamma_1$, and $\lambda$ can be
expressed through more commonly used Leslie viscosity coefficients
\cite{Forster:1971-1016,Pleiner:2002-297}. Note that equating
${\eta\equiv\nu_1=\nu_2=\nu_3}$ recovers the viscous stress tensor
$2\eta\mathbf S$ of the isotropic incompressible fluid (last term in
Eq.~(\ref{eq:SigmaI})).


\section{\label{sec:Dispersion}Dispersion relation}

The aim of this Section is to construct the dispersion relation for the surface
waves on the basis of the model set up above. We consider a surface wave with
frequency $\omega$ and wavevector $\mathbf q=(q,0,0)$ propagating along $x$
axis, and solve force balance equation, Eq.~(\ref{eq:forceBalance}), and torque
balance equation, Eq.~(\ref{eq:torqueBalance}) using linearized form of the
hydrodynamic equations written in Section~\ref{sec:Model}.

In order to linearize the hydrodynamic equations, we represent pressure
$p=p(\mathbf r,t)$ and the nematic director $\mathbf n=\mathbf n(\mathbf r,t)$,
where $\mathbf r=(x,y,z)$ is the position in space, $t$ is time, in form
\begin{equation}
p=p_0+\delta p,
\end{equation}
\begin{equation}
\label{eq:n}
\mathbf n=\mathbf n_0+\delta\mathbf n,
\end{equation}
where $\delta p$ and $\delta\mathbf n$ are the deviations of pressure and
director from their equilibrium values $p_0$ and $\mathbf n_0$, correspondingly.
The velocity $\mathbf v=\mathbf v(\mathbf r,t)$ is itself the deviation from
zero equilibrium velocity. Homeotropic anchoring corresponds to
\begin{equation}
\label{eq:n0}
\mathbf n_0=\left(0,0,1\right).
\end{equation}

For small deviations from the equilibrium, we shall use the hydrodynamic
equations linearized in $\mathbf v$, $\delta p$, and $\delta\mathbf n$. We shall
assume these quantities to be independent of the coordinate $y$
($\partial_y\equiv\partial/\partial y=0$) and vanish at $z\rightarrow\pm\infty$.

The magnetic field can be also represented as
$\boldsymbol{\mathcal H}=\boldsymbol{\mathcal H}_0+\delta\boldsymbol{\mathcal H}$,
where $\boldsymbol{\mathcal H}_0=\left(0,0,\mathcal H_0\right)$ is the
equilibrium value, and the deviation $\delta\boldsymbol{\mathcal H}$ can be
found from the linearized form of the Maxwell equations (\ref{eq:rotMaxwell}),
(\ref{eq:divMaxwell}). The terms in the final equations, containing
$\delta\boldsymbol{\mathcal H}$, are of higher order than linear, so we shall
use only the equilibrium value, and skip the `0' subscript, so that
$\boldsymbol{\mathcal H}=\left(0,0,\mathcal H\right)$.

Substituting the interfacial free energy density (\ref{eq:FS}) into
Eqs~(\ref{eq:def-tau}), (\ref{eq:def-xi}), (\ref{eq:def-M}), and
(\ref{eq:def-hSe}), we find the contributions up to the first order in
$\mathbf u$ (and its derivatives) and $\mathbf n_\Vert$ into surface tension
\begin{equation}
\label{eq:expl-tau}
\sigma=\bar\sigma+\frac W2\mathbf n_\Vert^2
-2\kappa\bar H^2+2\kappa\left(H-\bar H\right)^2+\bar\kappa K,
\end{equation}
bending moment tensor
\begin{equation}
\label{eq:expl-M}
\mathbf M=2\left[\left(\kappa+\bar\kappa\right)H-\kappa\bar H\right]\mathbf I_s
-\bar\kappa\mathbf b,
\end{equation}
tangential component of the capillary vector
\begin{equation}
\label{eq:expl-xi}
\boldsymbol\xi_\Vert=-W\mathbf n_\Vert,
\end{equation}
and tangential surface molecular field
\begin{equation}
\label{eq:expl-hS}
\mathbf h^{Sv}_\Vert=W\mathbf n_\Vert
-2\nabla_s\left[\left(\kappa+\bar\kappa\right)H-\kappa\bar H\right]
+\nabla_s\cdot\left(\bar\kappa\mathbf b\right).
\end{equation}
The non-vanishing components of the surface viscous stress tensor (\label{eq:sigmaSv}) are
\begin{equation}
\label{eq:expl-sigmaSv}
\boldsymbol\Sigma^{Sv}=2\eta_s\mathbf S^S
+\left(\zeta_s-\eta_s\right)\left(\mathbf I_s:\mathbf S^S\right)\mathbf I_s,
\end{equation}
The total interfacial force $\mathbf F^S$ can be found by substituting
Eqs~(\ref{eq:sigmaS}), (\ref{eq:sigmaSe}), and
(\ref{eq:expl-tau})--(\ref{eq:expl-sigmaSv}) into Eq.~(\ref{eq:def-ForceS}), and has
components
\begin{eqnarray}
\label{eq:FSx}
F^S_x&=&\epsilon_0\partial_x^2u_x+\left(\eta_s+\zeta_s\right)\partial_x^2v_x,
\\
\label{eq:FSy}
F^S_y&=&\eta_s\partial_x^2v_y,
\\
\nonumber
F^S_z&=&\left(\sigma_0+W_0\right)\partial_x^2u_z+W_0\partial_x\delta n_x
\\
\label{eq:FSz}
&+&\psi\partial_x^3u_x
-\kappa_0\partial_x^4u_z,
\end{eqnarray}
where
\begin{equation}
\psi\equiv2\left(\kappa_0\bar H_1+\kappa_1\bar H_0\right).
\end{equation}
To write the explicit form of the force balance equations
(\ref{eq:forceBalance}), we also need the expressions for the components of the
force (\ref{eq:def-ForceI}) exerted by the isotropic fluid,
\begin{eqnarray}
F^I_x&=&\eta\left(\partial_xv_z+\partial_zv_x\right)_{z=+0},
\\
F^I_y&=&\eta\left(\partial_zv_y\right)_{z=+0},
\\
F^I_z&=&\left(2\eta\partial_zv_z-p\right)_{z=+0},
\end{eqnarray}
and the components of the force (\ref{eq:def-ForceS}) exerted by the nematic
liquid crystal,
\begin{eqnarray}
F^N_x&=&\left[
\frac{1+\lambda}2h_x-\nu_3\left(\partial_xv_z+\partial_zv_x\right)
\right]_{z=-0},
\\
F^N_y&=&\left[\frac{1+\lambda}2h_y-\nu_3\partial_zv_y\right]_{z=-0},
\\
F^N_z&=&\left(p-2\nu_1\partial_zv_z\right)_{z=-0},
\end{eqnarray}
The hydrodynamic fields $\mathbf v$, $p$, $\mathbf n$ in the bulk isotropic and
nematic liquids are found by solution of the hydrodynamic expressions. The
explicit formulas are presented in Appendices~\ref{app:Isotropic} and
\ref{app:Nematic}.

Next we introduce Fourier transforms in the $x$ coordinate and in time as
\begin{equation}
\label{eq:vfourier}
\mathbf v\left(\mathbf r,t\right)=\frac1{\left(2\pi\right)^2}
\int\limits_{-\infty}^\infty dq\int\limits_{-\infty}^\infty d\omega
e^{i\omega t-iqx}\tilde{\mathbf v}\left(q,z,\omega\right),
\end{equation}
\begin{equation}
\label{eq:pfourier}
\delta p\left(\mathbf r,t\right)=\frac1{\left(2\pi\right)^2}
\int\limits_{-\infty}^\infty dq\int\limits_{-\infty}^\infty d\omega
e^{i\omega t-iqx}\tilde p\left(q,z,\omega\right),
\end{equation}
\begin{equation}
\label{eq:nfourier}
\delta\mathbf n\left(\mathbf r,t\right)=\frac1{\left(2\pi\right)^2}
\int\limits_{-\infty}^\infty dq\int\limits_{-\infty}^\infty d\omega
e^{i\omega t-iqx}\tilde{\mathbf n}\left(q,z,\omega\right)
\end{equation}
(for brevity we shall henceforth omit arguments of the transformed functions).
Performing Fourier-transform of the force balance equation
(\ref{eq:forceBalance}), and substituting
${\tilde{\mathbf u}=\tilde{\mathbf v}/i\omega}$, we obtain balance equations for
the force components in form
\begin{eqnarray}
\nonumber
-\epsilon^\ast_0 q^2\tilde v^S_x
-i\omega\eta\left(m^I+q\right)\tilde v^S_x
-\omega\eta\left(m^I-q\right)\tilde v^S_z-
\\
\nonumber
{}-\frac\omega q\nu_3
\sum_{i=1}^3\left[\left(m_i^{N\Vert}\right)^2+q^2\right]C_i^{N\Vert}+
\\
\nonumber
+i\omega\frac{1+\lambda}2
\sum_{i=1}^3\left[K_3\left(m_i^{N\Vert}\right)^2-K_1q^2+\chi_a\mathcal H^2\right]
B_i^\Vert C_i^{N\Vert}
\\
\label{eq:balanceVX}
=0,\quad
\end{eqnarray}
\begin{eqnarray}
\nonumber
-\eta_sq^2\tilde v^S_y
-\eta m^I\tilde v^S_y
-\nu_3\sum_{i=1}^2m_i^{N\bot}C_i^{N\bot}
+\frac{1+\lambda}2
\\
\nonumber
{}\times\sum_{i=1}^2\left[K_3\left(m_i^{N\bot}\right)^2-K_2q^2+\chi_a\mathcal H^2\right]
B_i^\bot C_i^{N\bot}
\\
\label{eq:balanceVY}
=0,
\end{eqnarray}
\begin{eqnarray}
\nonumber
-\left(\sigma_0+W_0\right)q^2\tilde v^S_z
+\omega qW_0\tilde n^S_x
-\kappa_0q^4\tilde v^S_z
&-&
\\
\nonumber
{}-iq^3\psi\tilde v^S_x
-2\omega q\eta\tilde v^S_x
+\frac{\omega^2\rho^I}q\frac{iq\tilde v^S_x+m^I\tilde v^S_z}{m^I-q}
&+&
\\
\label{eq:balanceVZ}
{}+i\omega\sum_{i=1}^3\left(A_i-2\nu_1m_i^{N\Vert}\right)C_i^{N\Vert}
&=&0,
\end{eqnarray}
where $\epsilon^\ast_0=\epsilon_0+i\omega\left(\eta_s+\zeta_s\right)$ is the
complex dilatational modulus, $m^I$ is defined in Appendix~\ref{app:Isotropic}
by Eq.~(\ref{eq:mI}), and the quantities $m^{N\Vert}_i$, $m^{N\bot}_i$,
$C^{N\Vert}_i$, $C^{N\bot}_i$, $B^{N\Vert}_i$, $B^{N\bot}_i$, and $A_i$ are
defined in Appendix~\ref{app:Nematic} by Eqs~(\ref{eq:mVert}), (\ref{eq:mbot}),
(\ref{eq:CVert}), (\ref{eq:Cbot}), (\ref{eq:BVert}), (\ref{eq:Bbot}),
and (\ref{eq:A}), correspondingly.

To write the interfacial torque balance equation (\ref{eq:torqueBalance}), we
cast the torque exerted upon the interface by the nematic liquid crystal,
$\mathbf T^N$, and the interfacial torque arising due to surface interactions,
$\mathbf T^S$, entering the interfacial torque balance equation
(\ref{eq:torqueBalance}), in form
\begin{equation}
\mathbf T^N=\mathbf k\times\mathbf h^N
\end{equation}
and
\begin{equation}
\mathbf T^S=\mathbf k\times\mathbf h^S,
\end{equation}
where the molecular field from the bulk
\begin{equation}
\mathbf h^N=
-\mathbf k\cdot
\left[\frac{\partial\mathcal F_F}{\partial\left(\nabla\mathbf n\right)}\right]_S
\end{equation}
has linearized components
\begin{equation}
h^N_x=-K_3\partial_z\delta n^S_x,
\end{equation}
\begin{equation}
h^N_y=-K_3\partial_z\delta n^S_y,
\end{equation}
and the surface molecular field $\mathbf h^S$ can be represented as a sum of
elastic ($\mathbf h^{Se}$) and viscous ($\mathbf h^{Sv}$) contributions
\begin{equation}
\mathbf h^S=\mathbf h^{Se}+\mathbf h^{Sv},
\end{equation}
given by Eqs~(\ref{eq:def-hSe}) and (\ref{eq:def-hSv}), correspondingly, and can
be represented in components as
\begin{equation}
\label{eq:TSx}
h^{Se}_x=W_0\left(\delta n^S_x+\partial_xu_z\right)
-\kappa_0\partial_x^3u_z
+\psi\partial_x^2u_x,
\end{equation}
\begin{equation}
\label{eq:TSy}
h^{Se}_y=W_0\delta n^S_y,
\end{equation}
\begin{equation}
h^{Sv}_x=\gamma^S_{1\Vert}
\frac{\partial\left(\delta n_x+\partial_xu_z\right)}{\partial t},
\end{equation}
\begin{equation}
h^{Sv}_y=\gamma^S_{1\Vert}\frac{\partial\delta n_y}{\partial t}.
\end{equation}
Then the surface torque balance equations can be written as
\begin{eqnarray}
\nonumber
{}-K_3\partial_z\delta n^S_x
+W_0\left(\delta n_x+\partial_xu_z\right)
-\kappa_0\partial_x^3u_z&+&
\\
{}+\psi\partial_x^2u_x+\gamma^S_{1\Vert}
\frac{\partial\left(\delta n_x+\partial_xu_z\right)}{\partial t}
&=&0,
\end{eqnarray}
\begin{equation}
{}-K_3\partial_z\delta n^S_y
+W_0\delta n_y
+\gamma^S_{1\Vert}\frac{\partial\delta n_y}{\partial t}=0,
\end{equation}
or, substituting the expressions (\ref{eq:solVZ}), (\ref{eq:solNX}) and
(\ref{eq:solNY}),
\begin{eqnarray}
\nonumber
\sum_{i=1}^3\left[\left(K_3m^{N\Vert}_i-W_0-i\omega\gamma^S_1\right)
B^\Vert_i\right.+
\\
\label{eq:balanceNX}
{}+\frac q\omega\left.\left(
W_0+\kappa_0q^2-\psi m^{N\Vert}_i+i\omega\gamma^S_1
\right)\right]C^{N\Vert}_i&=&0,
\end{eqnarray}
\begin{equation}
\label{eq:balanceNY}
\sum_{i=1}^2
\left(K_3m^{N\bot}_i-W_0-i\omega\gamma^S_1\right)B^\bot_iC^{N\bot}_i=0.
\end{equation}

The interfacial force balance equations
(Eqs~(\ref{eq:balanceVX})--(\ref{eq:balanceVZ}))
and the interfacial torque balance equations
(Eqs~(\ref{eq:balanceNX})--(\ref{eq:balanceNY})) form, with account of
Eqs~(\ref{eq:Cbot}) and (\ref{eq:CVert}), a homogeneous system of linear
algebraic equations in $\tilde v^S_x$, $\tilde v^S_y$, $\tilde v^S_z$,
$\tilde n^S_x$ and $\tilde n^S_y$. The dispersion relation is obtained from the
condition of existence of a solution to these equations, i.e. the requirement
for the determinant $D\left(\omega,q\right)$ of the matrix of coefficients for
this system to be zero
\begin{equation}
\label{eq:D}
D\left(\omega,q\right)=0.
\end{equation}

The equations (\ref{eq:balanceVY}) and (\ref{eq:balanceNY}) in
$\tilde v^S_y$ and $\tilde n^S_y$ decouple from the equations
(\ref{eq:balanceVX}), (\ref{eq:balanceVZ}) and (\ref{eq:balanceNX}) in
$\tilde v^S_x$, $\tilde v^S_z$ and $\tilde n^S_z$. Therefore, the matrix of
coefficients is block-diagonal, and the dispersion relation (Eq.~(\ref{eq:D}))
is equivalent to a pair of relations for $\left(x,z\right)$ and $y$ directions:
\begin{equation}
\label{eq:DVert}
D_\Vert\left(\omega,q\right)=0,
\end{equation}
\begin{equation}
\label{eq:Dbot}
D_\bot\left(\omega,q\right)=0,
\end{equation}
where $D_\Vert\left(\omega,q\right)$ is the determinant of the ${3\times3}$
matrix $M_\Vert$ of coefficients for the equations (\ref{eq:balanceVX}),
(\ref{eq:balanceVZ}) and (\ref{eq:balanceNX}), and $D_\bot\left(\omega,q\right)$
is the determinant of the ${2\times2}$ matrix of coefficients for the equations
(\ref{eq:balanceVY}) and (\ref{eq:balanceNY}).

The explicit form of the dispersion relations is presented in
Appendix~\ref{app:Dispersion} and can be readily used for the numerical analysis
of surface modes.


\section{\label{sec:Modes}Surface modes}

In this Section the dispersion equation, which is presented in
Appendix~\ref{app:Dispersion}, is solved numerically, and surface modes of
different types are analyzed. For simplicity, we assume the density of the
isotropic liquid, $\rho^I$, to be small enough to be neglected
(e.g.\ nematic--surfactant--air interface). We also assume that the magnetic
field is absent.

The surface modes can be easily classified at low wavevectors $q$. Expansion of
the dispersion relation in powers of the wavevector $q$ is a straightforward
exercise in algebra, and the resulting modes are described below.

Firstly, there is a transverse capillary mode, which has the dispersion law
similar to that in the case of an isotropic liquid-liquid interface
\cite{Kats:1988-940,Kramer:1971-2097}:
\begin{equation}
\label{eq:omegaC}
\omega_C\left(q\right)=
\sqrt{\frac{\sigma_0q^3}{\rho^N}}+o\left(q^{3/2}\right).
\end{equation}
The principal contribution to this mode at large wavelengths arises due to the
restoring influence of surface tension $\sigma_0$, and the predominant motion is
in the direction normal to the interface ($z$). The differences from the isotropic
case, related to anisotropy of viscous dissipation in the nematic, appear in
higher orders in $q$.

The dilatational (or compressional) mode with predominant motion in the
direction along wave propagation ($x$) arises in presence of surfactant layer
due to the restoring force provided by the dilatational elastic modulus
$\epsilon_0$. The dispersion law for this mode can be written as
\begin{equation}
\label{eq:omegaD}
\omega_D\left(q\right)=
\left[\frac{i\epsilon_0^2q^4\nu_3}{\rho^N\left(\eta_2^M\right)^2}\right]^{1/3}
+o\left(q^{4/3}\right),
\end{equation}
where the Miesowicz viscosity $\eta_2^M$ is given by \cite{Forster:1971-1016}
\begin{equation}
\eta_2^M=\nu_3+\frac{\left(1+\lambda\right)^2}4\gamma_1.
\end{equation}
The difference from the dispersion law for the dilatational mode in the case of
a surfactant film at the interface between isotropic fluids, given by
\cite{Kats:1988-940,Kramer:1971-2097}
\begin{equation}
\label{eq:omegaI}
\omega^I_D\left(q\right)=
\left(\frac{i\epsilon_0^2q^4}{\rho^I\eta}\right)^{1/3}
+o\left(q^{4/3}\right),
\end{equation}
arises due to anisotropy of viscous dissipation in nematic.

A new mode, specific to the nematic, is driven by relaxation of the director
field to equilibrium due to anchoring at the interface and has the disperion law
\begin{equation}
\label{eq:omegaN}
\omega_N\left(q\right)=\frac{iW_0}{\gamma^S_1}+o\left(q^0\right).
\end{equation}
Such relaxation is present even in absence of motion of the interface
(e.g.\ when the interface is solid), so that $i\omega_N$ does not vanish at
$q\rightarrow0$. For nematic-isotropic interfaces, the
corresponding motion of the interface is induced by backflow effects.

Finally, behavior of the in-plane shear mode, with motion in $y$-direction,
is also governed by relaxation of the nematic director due to anchoring. The
corresponding dispersion law
\begin{equation}
\label{eq:omegaS}
\omega_S\left(q\right)=\frac{iW_0}{\gamma^S_1}+o\left(q^0\right)
\end{equation}
appears to be different from the isotropic case, where the damping of the
in-plane shear mode in absence of anchoring is governed by the surface viscosity
$\eta_s$ \cite{Kramer:1971-2097}.

Gravity $g$, so far neglected in our analysis, becomes important at wavevectors
\begin{equation}
\label{eq:qg}
q\sim q_g=\sqrt{\frac{\left|\rho^N-\rho^I\right|g}{\sigma_0+W_0}},
\end{equation}
and can be taken into account by adding the hydrostatic pressure term
$-g\left|\rho^N-\rho^I\right|$ to Eq.~(\ref{eq:M22}), which corresponds to the
additional contribution $-g\left|\rho^N-\rho^I\right|u_z$ to the vertical
component of the force, Eq.~(\ref{eq:FSz}). The resulting dispersion law for
transversal mode is given by expression
\begin{equation}
\label{eq:omegaG}
\omega_G\left(q\right)=\sqrt{gq}+o\left(q^{1/2}\right),
\end{equation}
which describes well-known gravity waves \cite{Landau:FM}.

In the opposite case of large wavevectors, the curvature energy becomes important.
Analysis of Eqs~(\ref{eq:FSz}) and (\ref{eq:TSx}) yields the characteristic
values of $q$
\begin{equation}
\label{eq:qkappa}
q\sim q_\kappa=\sqrt{\frac{W_0}{\kappa_0}}
\end{equation}
and
\begin{equation}
q\sim q_\psi=\frac{W_0}{\left|\kappa_0\bar H_1+\kappa_1\bar H_0\right|}
\end{equation}
below which one can neglect in the dispersion relation the terms containing
bending rigidity $\kappa$ and its derivatives with respect to surfactant
concentration, given by $\psi$.

Usually $q_\psi>q_\kappa$, and the range of $q$ in which both gravity and
curvature contributions become small, given by
\begin{equation}
\label{eq:good-q}
q_g\ll q\ll q_\kappa,
\end{equation}
is rather wide. For typical values
$\left|\rho^N-\rho^I\right|\sim10$\,kg/m$^3$,
$g\sim10$\,m/s$^2$,
$\sigma_0\sim W_0\sim10^{-2}$\,J/m$^2$,
$\kappa_0\sim10^{-18}$\,J,
the equation (\ref{eq:good-q}) reads
$1$\,cm$^{-1}\ll q\ll10^6$\,cm$^{-1}$,
which includes the range of wavevectors typically probed by surface light
scattering experiments.

To obtain the dispersion laws for surface modes at larger values of the
wavevector $q$, the dispersion equation must be solved numerically. The
numerical solution presented below uses the following typical values of the
material parameters when it is not indicated otherwise.
For the nematic liquid crystal we use the parameters of
4-$n$-pentyl-4'-cyanobiphenyl
(5CB) at 26$^o$C \cite{Borzsonyi:1998-7419}:
the density
$\rho^N=1021.5$\,kg/m${}^3$,
the elastic constants
$K_1=5.95\cdot10^{-12}$\,N,
$K_2=3.77\cdot10^{-12}$\,N,
$K_3=7.86\cdot10^{-12}$\,N,
the Leslie viscosities
$\alpha_1=-6.6\cdot10^{-3}$\,kg/(m$\cdot$s),
$\alpha_2=-77.0\cdot10^{-3}$\,kg/(m$\cdot$s),
$\alpha_3=-4.2\cdot10^{-3}$\,kg/(m$\cdot$s),
$\alpha_4=63.4\cdot10^{-3}$\,kg/(m$\cdot$s),
$\alpha_5=62.4\cdot10^{-3}$\,kg/(m$\cdot$s),
$\alpha_6=-18.4\cdot10^{-3}$\,kg/(m$\cdot$s).
The viscosity coefficient used in the present paper can be calculated
from the Leslie equations \cite{Forster:1971-1016,Pleiner:2002-297} and equal
$\nu_1=50.4\cdot10^{-3}$\,kg/(m$\cdot$s),
$\nu_2=31.7\cdot10^{-3}$\,kg/(m$\cdot$s),
$\nu_3=19.96\cdot10^{-3}$\,kg/(m$\cdot$s),
$\gamma_1=72.8\cdot10^{-3}$\,kg/(m$\cdot$s),
$\lambda=1.115$.
We use the value of the bending rigidity
$\kappa_0=10^{-19}$\,J
which is typical for surfactant layers \cite{Marsh:2006-146}.
For other parameters we use the following typical values:
$\gamma^S_1\sim5\cdot10^{-9}$\,kg/s,
$\bar H_0=0$,
$\epsilon_0=10^{-3}$\,N/m,
$\eta_s=10^{-8}$\,N/m,
$\sigma_0=5\cdot10^{-3}$\,J/m${}^2$,
$W_0=20\cdot10^{-3}$\,J/m${}^2$.

\begin{figure}
\begin{center}
\includegraphics[width=1.0\columnwidth,keepaspectratio]{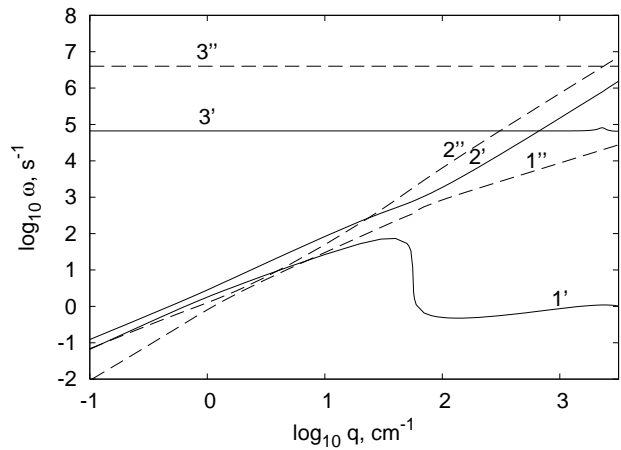}
\end{center}
\caption{
\label{fig:nogravity}
Dispersion law $\omega\left(q\right)$ for different surface modes in absence of
gravity, obtained by solution of the dispersion relation (\ref{eq:DVert}) with
the values of the parameters given in the text. Numbers 1, 2, 3 denote
transverse, dilatational, and nematic director relaxation modes,
correspondingly. Prime and double prime denote real (solid line) and imaginary
(dashed line) parts of $\omega$, correspondingly.
}
\end{figure}

The dispersion law $\omega\left(q\right)$ for different surface modes in absence
of gravity, obtained by solution of the dispersion relation (\ref{eq:DVert})
with the values of the parameters given above, is presented in
Figure~\ref{fig:nogravity}. At low $q$ the dispersion of for modes 1, 2, 3, as
denoted Figure~\ref{fig:nogravity}, is in good agreement with approximate
formulas (\ref{eq:omegaC}), (\ref{eq:omegaD}), and (\ref{eq:omegaN}),
correspondingly. The noticeable discrepancy in behavior of capillary and
dilatational modes appears at $q\sim10$\,cm$^{-1}$, and the damping of surface
waves becomes large at larger $q$, which is qualitatively similar to the case of
the interface between isotropic liquids.

The results presented in Figure~\ref{fig:nogravity} suggest that in the
typical range of $q$ probed by surface light scattering experiments
($100$\,cm$^{-1}\lesssim q\lesssim2000$\,cm$^{-1}$), the approximate expressions
(\ref{eq:omegaC}), (\ref{eq:omegaD}) do not describe well the dispersion curves,
and accurate solution of the dispersion equation should be used instead.

\begin{figure}
\begin{center}
\includegraphics[width=1.0\columnwidth,keepaspectratio]{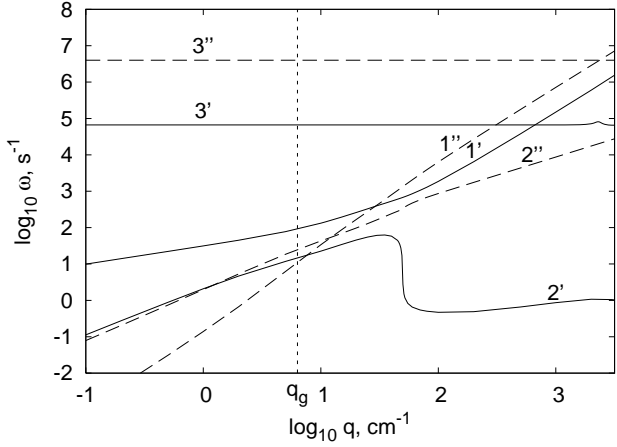}
\end{center}
\caption{
\label{fig:gravity}
Dispersion law $\omega\left(q\right)$ for different surface modes in presence of
gravity $g=9.8$\,m/s$^2$, obtained by solution of the dispersion relation
(\ref{eq:DVert}) with the values of the parameters given in the text. Numbers
1, 2, 3 denote transverse, dilatational, and nematic director relaxation modes,
correspondingly. Prime and double prime denote real (solid line) and imaginary
(dashed line) parts of $\omega$, correspondingly. Vertical dotted line
corresponds to the value of $q_g$ given by Eq.~(\ref{eq:qg}).
}
\end{figure}

Figure~\ref{fig:gravity} presents the dispersion law $\omega\left(q\right)$ for
different surface modes obtained by solution of the dispersion relation
(\ref{eq:DVert}) in presence of gravity $g=9.8$\,m/s$^2$. In agreement with the
discussion above, the influence of gravity on the dispersion laws is small at
$q\gg q_g$, where $q_g$ is given by Eq.~(\ref{eq:qg}).

\begin{figure}
\begin{center}
\includegraphics[width=1.0\columnwidth,keepaspectratio]{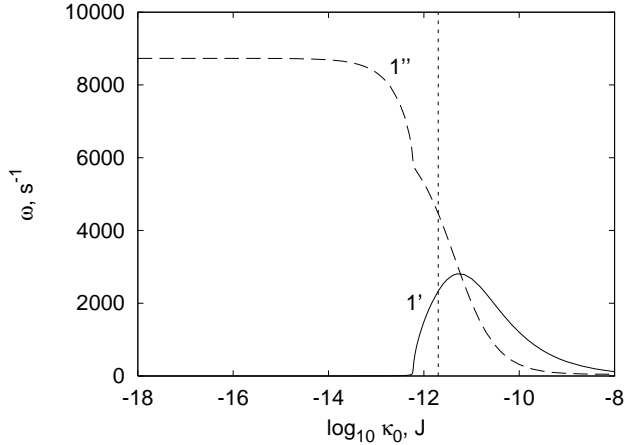}
\end{center}
\caption{
\label{fig:bending}
Dependence of the real (solid line) and imaginary (dashed line) parts of the
frequency of the mode 1 (as defined on Figure~\ref{fig:nogravity}) upon the
bending rigidity $\kappa_0$, calculated at $q=1000$\,cm$^{-1}$ in absence of
gravity. Vertical line corresponds to the value of $\kappa_0$ that satisfies
Eq.~(\ref{eq:qkappa}).
}
\end{figure}

If the bending rigidity $\kappa$ is large, its influence becomes noticeable, as
it is demonstrated in Figure~\ref{fig:bending}. For $\kappa\sim kT$, typical
for surfactant films, the value of $q_\kappa$, given by eq.~(\ref{eq:qkappa}),
corresponds to wavelength close to atomic scales, and curvature energy can be
neglected in typical surface light scattering experiments, in agreement with the
discussion above.

\begin{figure}
\begin{center}
\includegraphics[width=1.0\columnwidth,keepaspectratio]{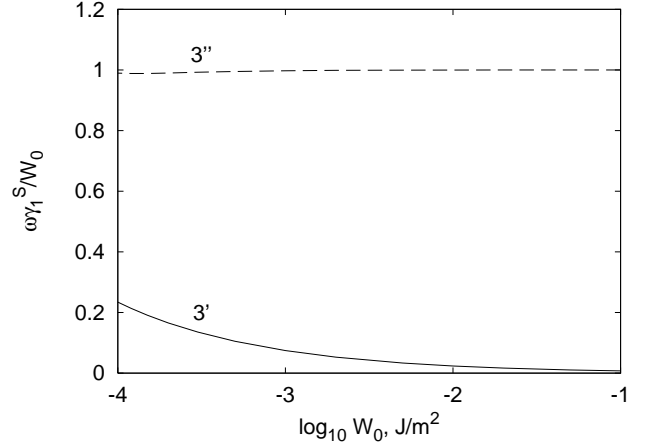}
\end{center}
\caption{
\label{fig:xanchoring}
Dependence of the real (solid line) and imaginary (dashed line) parts of the
frequency $\omega$ of the mode 3 (as defined on Figure~\ref{fig:nogravity}),
normalized by $W_0/\gamma^S_1$ (see Eq.~(\ref{eq:omegaN})), upon the anchoring
strength $W_0$, calculated at $q=100$\,cm$^{-1}$ in absence of gravity.
}
\end{figure}

\begin{figure}
\begin{center}
\includegraphics[width=1.0\columnwidth,keepaspectratio]{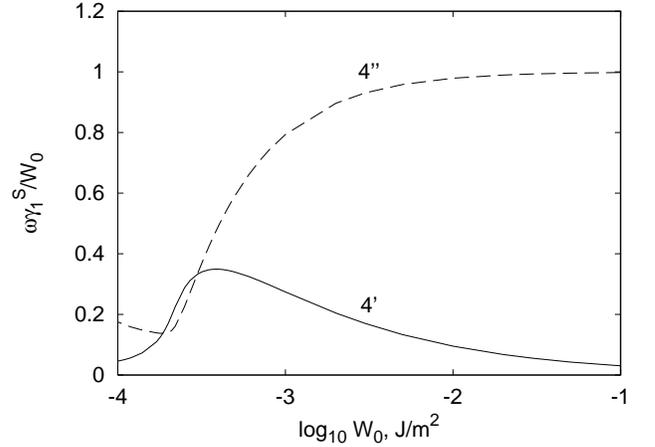}
\end{center}
\caption{
\label{fig:yanchoring}
Dependence of the real (solid line) and imaginary (dashed line) parts of the
frequency $\omega$ of the in-plane shear mode, normalized by
$W_0/\gamma^S_1$ (see Eq.~(\ref{eq:omegaS})), upon the anchoring
strength $W_0$, calculated at $q=100$\,cm$^{-1}$ in absence of gravity.
}
\end{figure}

The dispersion law for the modes governed by relaxation of the nematic director
field in $x$ and $y$ directions due to anchoring of the nematic director at the
interface, obtained by numerical solution of the dispersion equation with the
values of the parameters given above, are well described by the equations
(\ref{eq:omegaN}) and (\ref{eq:omegaS}). However, as the anchoring strength
becomes smaller, other mechanisms start to take over, as demonstrated in
Figures~\ref{fig:xanchoring} and \ref{fig:yanchoring}.


\section{\label{sec:Field}Influence of magnetic field}

In this Section we discuss how the surface modes described in
Section~\ref{sec:Modes} are altered in presence of the external magnetic field
directed normally to the surface (along $z$ axis).

The external magnetic field effectively acts on the nematic molecules as
an additional molecular field (see Eq.~(\ref{eq:h-star})), and the primary
counteracting mechanism is provided by orientational shear relaxation. Thus we
may expect the influence of the magnetic field become noticeable at
\begin{equation}
\label{eq:Hstar}
\chi_a\mathcal H^2\sim\omega\gamma_1.
\end{equation}

\begin{figure}
\begin{center}
\includegraphics[width=1.0\columnwidth,keepaspectratio]{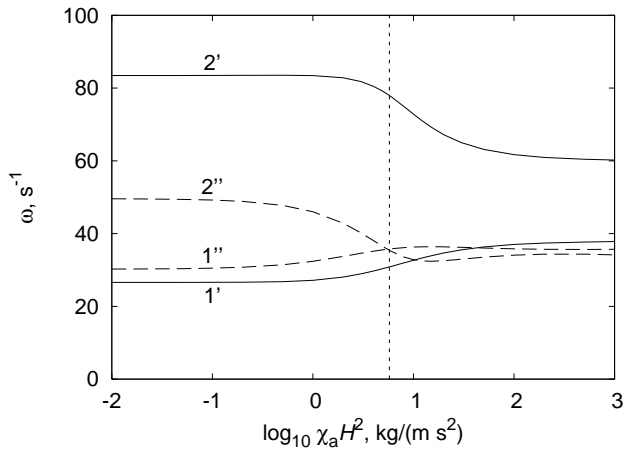}
\end{center}
\caption{
\label{fig:field}
Dependence of the real (solid lines) and imaginary (dashed lines) parts of the
frequencies of the modes 1 and 2 (as defined on Figure~\ref{fig:nogravity}) upon
the magnetic field, calculated at $q=10$\,cm$^{-1}$ in absence of gravity.
Vertical dotted line corresponds to $\chi_a\mathcal H^2=\omega\gamma_1$ (see
Eq.~(\ref{eq:Hstar})).
}
\end{figure}

The results of the numerical solution of the dispersion equation in presence of
magnetic field, presented in Figure~\ref{fig:field}, confirm that noticeable
change in dispersion of capillary and dilatational modes arises only around the
value of the field given by Eq.~(\ref{eq:Hstar}). The change due to magnetic
field in modes governed by anchoring is found to be negligibly small.

At low $q$ the dispersion of a capillary mode in strong magnetic field is
different from the law (\ref{eq:omegaC}) and is given by
\begin{equation}
\label{eq:omegaM}
\omega_C=
\sqrt{\frac{\left(\sigma_0+W_0\right)q^3}{\rho^N}}+o\left(q^{3/2}\right).
\end{equation}
The frequency of this mode becomes sensitive to the anchoring properties of the
interface, because the nematic director tends to be oriented along the field
rather than to be advected with the nematic liquid. The practical use of this
effect is, however, limited, because at short wavelengths extremely large
magnetic field is required, and at long wavelengths gravity becomes dominating
(Eq.~(\ref{eq:omegaG}).

In principle, magnetic field can also influence surface waves through change in
the properties of the interface (e.g.\ surface tension) due to the magnetization
of the surfactant. Separate study is required to estimate the magnitude of this
effect.


\section{\label{sec:Conclusion}Conclusion}

We have obtained the dispersion relation for the surface waves at a
surfactant-laden nematic isotropic interface for the case when the surfactant
film induces homeotropic (normal to the surface) orientation of the director,
and the surfactant film is in the isotropic two-dimensional fluid phase. We have
analyzed the dispersion law of different surface modes analytically in long
wavelength limit, and numerically in broader range of wave vectors, using
typical values of the material parameters.

At long wavelengths the dispersion of capillary, dilatational (or compression),
in-plane shear, and director relaxation modes is described by equations
(\ref{eq:omegaC}) (or (\ref{eq:omegaG})), (\ref{eq:omegaD}), (\ref{eq:omegaS}),
and (\ref{eq:omegaN}), correspondingly. At smaller wavelength, the solution of
the full dispersion relation should be used. Gravity influences the transversal
mode at small wavevectors (Eq.~(\ref{eq:qg})), and curvature energy of
surfactant can be neglected if wavevector is not too large
(Eq.~(\ref{eq:qkappa})). For all modes, the influence of the external magnetic
field directed normally to the interface is small.

The influence of the magnetic field should be more pronounced if the direction
of the field does not coincide with equilibrium nematic director. In this case
the dispersion law for surface modes may be expected to be quantitatively
different due to anisotropy of viscous dissipation in nematic, and different
anchoring energy. The results of the present paper can be readily extended to
the case of arbitrary direction of the external field and to other types of
nematic anchoring.

Other possible developments, which may increase the range of accessible systems
and conditions, is the extension of the results to wider range of the states of
the surfactant film, and the study of the effects which may be caused by the
phase transitions in the surfactant film and bulk liquid crystal.

Dependence of the dispersion waves upon the parameters of the interface suggests
the surface light scattering on a surfactant-laden nematic-isotropic interface
as a potential method for determining of the properties of surfactant-laden
nematic-isotropic interfaces, and as a possible candidate for a chemical or
biological sensing technique.


\acknowledgments

I thank Prof.\ C.~M.~Care for fruitful discussion of the results, and
Prof.\ P.~D.~I.~Fletcher for the discussion about surfactant-laden
nematic-isotropic interfaces which instigated this work.

\appendix


\section{\label{app:Geometry}Differential geometry of the interface}

The geometrical description we use is similar to that of that presented in works
\cite{Buzza:2002-8418} and \cite{Rey:2006-219}. We choose the plane $z=0$ to
coincide with the unperturbed interface, the half-space $z<0$ to be occupied by
the uniaxial nematic liquid crystal, and the half-space $z>0$ to be filled by
the isotropic liquid.

Let the position of a fluid particle at the interface be
$\mathbf r=\mathbf r_0+\mathbf u$, where $\mathbf r_0=\left(x_0,y_0,0\right)$ is
its position on the undeformed interface ($z=0$), and
$\mathbf u=\mathbf u\left(\mathbf r_0\right)$ is the displacement vector with
components $\left(u_x,u_y,u_z\right)$. We shall use $x_0$ and $y_0$ as surface
coordinates and denote them as $s^\alpha$, $\alpha$ and other Greek indices
taking values 1 and 2.

The position $\mathbf r$ of fluid particles at the interface in 3D space can be
cast as
\begin{equation}
\mathbf r=\mathbf R(s^\alpha).
\end{equation}
The surface tangent base vectors
$\mathbf a_\alpha=\partial\mathbf r/\partial s^\alpha$, corresponding to the
chosen surface coordinates, can be written in terms of the components of the
displacement vectors:
\begin{equation}
\mathbf a_1=\frac{\partial\mathbf r}{\partial s^1}=
\begin{pmatrix}
1+\partial_xu_x, & \partial_xu_y, & \partial_xu_z
\end{pmatrix}
\end{equation}
and
\begin{equation}
\mathbf a_2=\frac{\partial\mathbf r}{\partial s^2}=
\begin{pmatrix}
\partial_yu_x, & 1+\partial_yu_y, & \partial_yu_z
\end{pmatrix}.
\end{equation}
The surface metric tensor
\begin{equation}
a_{\alpha\beta}=\mathbf a_\alpha\cdot\mathbf a_\beta=
\begin{pmatrix}
1+2\partial_xu_x & \partial_xu_y+\partial_yu_x
\\
\partial_xu_y+\partial_yu_x & 1+2\partial_yu_y
\end{pmatrix}
+O(u^2),
\end{equation}
has determinant
\begin{equation}
a=\det(a_{\alpha\beta})
=1+2\left(\partial_xu_x+\partial_yu_y\right)
+O(u^2).
\end{equation}
The corresponding reciprocal base vectors $\mathbf a^\alpha$ and metric tensor
$a^{\alpha\beta}$ take form
\begin{equation}
\mathbf a^1=\frac{\partial s^1}{\partial\mathbf r}=
\begin{pmatrix}
1-\partial_xu_x, & -\partial_yu_x, & \partial_xu_z
\end{pmatrix}
+O(u^2),
\end{equation}
\begin{equation}
\mathbf a^2=\frac{\partial s^2}{\partial\mathbf r}=
\begin{pmatrix}
-\partial_xu_y, & 1-\partial_yu_y, & \partial_yu_z
\end{pmatrix}
+O(u^2),
\end{equation}
\begin{eqnarray}
a^{\alpha\beta}&=&\mathbf a^\alpha\cdot\mathbf a^\beta=
\\
\nonumber
&=&
\begin{pmatrix}
1-2\partial_xu_x & -\partial_xu_y-\partial_yu_x
\\
-\partial_xu_y-\partial_yu_x & 1-2\partial_yu_y
\end{pmatrix}
+O(u^2).
\end{eqnarray}
The base and reciprocal base vectors satisfy
\begin{equation}
\mathbf a_\alpha\cdot\mathbf a^\beta=\delta_\alpha^\beta=
\begin{pmatrix}
1 & 0 \\ 0 & 1
\end{pmatrix},
\end{equation}

We write the unit vector $\mathbf k$, normal to the interface and directed into
the isotropic liquid, as
\begin{equation}
\mathbf k=\frac{\mathbf a_1\times\mathbf a_2}{|\mathbf a_1\times\mathbf a_2|}=
\begin{pmatrix}
-\partial_xu_z, & -\partial_yu_z, & 1
\end{pmatrix}
+O(u^2).
\end{equation}
We shall also define the dyadic surface idem factor
\begin{equation}
\mathbf I_s=\mathbf a_\alpha\mathbf a^\alpha=
\begin{pmatrix}
1 & 0 & \partial_xu_z
\\
0 & 1 & \partial_yu_z
\\
\partial_xu_z & \partial_yu_z & 0
\end{pmatrix}
+O(u^2).
\end{equation}
the surface gradient operator
\begin{equation}
\nabla_s=\mathbf I_s\cdot\nabla=
\begin{pmatrix}
\partial_x+\left(\partial_xu_z\right)\partial_z
\\
\partial_y+\left(\partial_yu_z\right)\partial_z
\\
\left(\partial_xu_z\right)\partial_x+\left(\partial_yu_z\right)\partial_y
\\
\end{pmatrix}
+O(u^2).
\end{equation}
and the second fundamental tensor
\begin{equation}
\mathbf b=-\nabla_s\mathbf k=
\begin{pmatrix}
\partial_x^2u_z & \partial_x\partial_yu_z & 0
\\
\partial_x\partial_yu_z & \partial_y^2u_z & 0
\\
0 & 0 & 0
\end{pmatrix}
+O(u^2).
\end{equation}
The mean curvature $H$ and Gaussian curvature $K$ are given by
\begin{equation}
H=\frac12(\mathbf I_s:\mathbf b)=\frac12\left(
\partial_x^2u_z+\partial_y^2u_z
\right)+O(u^2),
\end{equation}
\begin{equation}
K=\frac12
\epsilon^{\alpha\beta}\epsilon^{\gamma\delta}b_{\alpha\gamma}b_{\beta\delta}
=O(u^2).
\end{equation}

Other useful identities include the surface projection $\mathbf n_\Vert$ of a
nematic director field $\mathbf n$, Eqs~(\ref{eq:n}) and (\ref{eq:n0}),
\begin{equation}
\mathbf n_\Vert=\mathbf I_S\cdot\mathbf n=
\begin{pmatrix}
\delta n_x+\partial_xu_z
\\
\delta n_y+\partial_yu_z
\\
0
\end{pmatrix}+o\left(\nabla\mathbf u,\delta\mathbf n\right),
\end{equation}
and its surface divergence
\begin{equation}
\nabla_S\cdot\mathbf n_\Vert=
\partial_x\delta n_x+\partial_y\delta n_y+\partial_x^2u_z+\partial_y^2u_z
+o\left(\nabla\mathbf u,\delta\mathbf n\right).
\end{equation}


\section{\label{app:Isotropic}Bulk solution for isotropic liquid}

This appendix presents the solution to the linearized hydrodynamic equations in
bulk isotropic liquid, obtained by Kramer \cite{Kramer:1971-2097}.

Substitution of Eq.~(\ref{eq:vfourier}) into Eq.~(\ref{eq:continuity}) yields
\begin{equation}
\label{eq:continuity-fourier}
-iq\tilde v_x+\partial_z\tilde v_z=0.
\end{equation}
Substituting Eqs~(\ref{eq:vfourier}) and (\ref{eq:pfourier}) into
Eqs~(\ref{eq:Navier-Stokes})--(\ref{eq:S}), we obtain
\begin{eqnarray}
\label{eq:Kramer-vx}
\left[i\omega\rho^I+\eta\left(q^2-\partial_z^2\right)\right]
\tilde v_x&=&iq\tilde p,
\\
\label{eq:Kramer-vy}
\left[i\omega\rho^I+\eta\left(q^2-\partial_z^2\right)\right]
\tilde v_y&=&0,
\\
\label{eq:Kramer-vz}
\left[i\omega\rho^I+\eta\left(q^2-\partial_z^2\right)\right]
\tilde v_z&=&-\partial_z\tilde p,
\end{eqnarray}
where equation (\ref{eq:Kramer-vy}) is decoupled from other equations. The
general solution to Eqs~(\ref{eq:continuity-fourier})--(\ref{eq:Kramer-vz})
vanishing at $z\rightarrow\infty$ can be written as
\begin{equation}
\tilde v_x=iC^{I\Vert}_1e^{-qz}+i\frac{m^I}qC^{I\Vert}_2e^{-m^Iz},
\end{equation}
\begin{equation}
\tilde v_y=C^{I\bot}e^{-m^Iz},
\end{equation}
\begin{equation}
\tilde v_z=C^{I\Vert}_1e^{-qz}+C^{I\Vert}_2e^{-m^Iz},
\end{equation}
\begin{equation}
\tilde p=\frac{i\omega\rho^I}qC^{I\Vert}_1e^{-qz},
\end{equation}
with
\begin{equation}
\label{eq:mI}
m^I=\left(q^2+\frac{i\omega\rho^I}\eta\right)^{1/2},\quad\textrm{Re}\,m^I>0.
\end{equation}
The quantities $C^{I\Vert}_1$, $C^{I\Vert}_2$, and $C^{I\bot}$ are functions of
$q$ and $\omega$ and are determined by the boundary conditions at the interface
as follows:
\begin{equation}
C^{I\Vert}_1=\frac{iq\tilde v_x^S+m^I\tilde v_z^S}{m^I-q},
\end{equation}
\begin{equation}
C^{I\bot}=\tilde v_y^S,
\end{equation}
\begin{equation}
C^{I\Vert}_2=-\frac{iq\tilde v_x^S+q\tilde v_z^S}{m^I-q},
\end{equation}
where the superscript $S$ indicates that the values of the corresponding dynamic
variables are taken at ${z\rightarrow0}$.


\section{\label{app:Nematic}Bulk solution for nematic liquid crystal}

In this appendix the solution is presented to the linearized hydrodynamic
equations in bulk nematic liquid crystal.

For the equilibrium director along $z$ axis (Eq.~(\ref{eq:n0})) the
Fourier-transform, similar to equations
(\ref{eq:vfourier})--(\ref{eq:nfourier}), of the linearized molecular field
(Eq.~(\ref{eq:h})), $\tilde{\mathbf h}$, has non-zero components
\begin{equation}
\label{eq:hx}
\tilde h_x=-K_1q^2\tilde n_x+K_3\partial_z^2\tilde n_x+\chi_a\mathcal H^2\tilde n_x,
\end{equation}
\begin{equation}
\label{eq:hy}
\tilde h_y=-K_2q^2\tilde n_y+K_3\partial_z^2\tilde n_y+\chi_a\mathcal H^2\tilde n_y.
\end{equation}
Substituting them into Eqs~(\ref{eq:EriksenLeslie}),
(\ref{eq:SigmaN})--(\ref{eq:SigmaNv}), we obtain the following linear
differential equations,
\begin{equation}
\label{eq:Kramer-vxN}
\left[i\omega\rho^N+\left(2\nu_2-\nu_3\right)q^2
-\nu_3\partial_z^2\right]\tilde v_x-\frac{1-\lambda}2\partial_z\tilde h_x
=iq\tilde p,
\end{equation}
\begin{equation}
\label{eq:Kramer-vyN}
\left[i\omega\rho^N+\nu_2q^2
-\nu_3\partial_z^2\right]\tilde v_y
-\frac{1-\lambda}2\partial_z\tilde h_y=0,
\end{equation}
\begin{equation}
\label{eq:Kramer-vzN}
\left[i\omega\rho^N+\nu_3q^2-\left(2\nu_1-\nu_3\right)\partial_z^2\right]
\tilde v_z-iq\frac{1+\lambda}2\tilde h_x=-\partial_z\tilde p,
\end{equation}
which are analogous to Eqs~(\ref{eq:Kramer-vx})--(\ref{eq:Kramer-vz}) for
isotropic liquids. Equation (\ref{eq:dndt}) for the director after Fourier
transform gives two equations,
\begin{equation}
\label{eq:Kramer-nxN}
i\omega\tilde n_x=\frac{1+\lambda}2\partial_z\tilde v_x
+\frac{1-\lambda}2iq\tilde v_z+\frac1{\gamma_1}\tilde h_x
\end{equation}
and
\begin{equation}
\label{eq:Kramer-nyN}
i\omega\tilde n_y=\frac{1+\lambda}2\partial_z\tilde v_y
+\frac1{\gamma_1}\tilde h_y,
\end{equation}
where $\tilde h_x$ and $\tilde h_y$ are given by Eqs~(\ref{eq:hx}) and
(\ref{eq:hy}).

Thus we have six linear differential equations
(Eqs~(\ref{eq:continuity-fourier}) and
(\ref{eq:Kramer-vxN})--(\ref{eq:Kramer-nyN})) for six dynamic variables
(pressure, three components of velocity, and two components of director).
Equations (\ref{eq:Kramer-vyN}) and (\ref{eq:Kramer-nyN}) for $\tilde v_y$ and
$\tilde n_y$ decouple from the others, their general solution vanishing at
${z\rightarrow-\infty}$ can be cast as
\begin{equation}
\label{eq:solVY}
\tilde v_y=\sum_{i=1}^2C^{N\bot}_ie^{m^{N\bot}_iz},
\end{equation}
\begin{equation}
\label{eq:solNY}
\tilde n_y=\sum_{i=1}^2B^\bot_iC^{N\bot}_ie^{m^{N\bot}_iz},
\end{equation}
where
\begin{equation}
\label{eq:Bbot}
B^\bot_i=\frac{\left(1+\lambda\right)\gamma_1m^{N\bot}_i}
{2\left[i\omega\gamma_1+K_2q^2-K_3\left(m^{N\bot}_i\right)^2-\chi_a\mathcal H^2\right]},
\end{equation}
\begin{equation}
\label{eq:mbot}
m^{N\bot}_i=\left(\mu^\bot_i\right)^{1/2},\quad\textrm{Re}\,m^{N\bot}_i>0,
\end{equation}
and $\mu^\bot_i$, $i=1,2$, are the roots of the quadratic equation
\begin{equation}
\label{eq:Quadratic}
a^\bot_2\left(\mu^\bot\right)^2+a^\bot_1\mu^\bot+a^\bot_0=0,
\end{equation}
where
\begin{equation}
a^\bot_2=\left(\frac{1-\lambda^2}4\gamma_1-\nu_3\right)K_3,
\end{equation}
\begin{eqnarray}
\nonumber
a^\bot_1&=&\left(i\omega\gamma_1+K_2q^2-\chi_a\mathcal H^2\right)\nu_3
+\left(i\omega\rho^N+\nu_2q^2\right)K_3
\\
&-&\frac{1-\lambda^2}4\gamma_1\left(K_2q^2-\chi_a\mathcal H^2\right),
\end{eqnarray}
\begin{equation}
a^\bot_0=-\left(i\omega\rho^N+\nu_2q^2\right)
\left(i\omega\gamma_1+K_2q^2-\chi_a\mathcal H^2\right).
\end{equation}

The general solution to the equations (\ref{eq:continuity-fourier}),
(\ref{eq:Kramer-vxN}), (\ref{eq:Kramer-vzN}), and (\ref{eq:Kramer-nxN})
vanishing at ${z\rightarrow-\infty}$ can be cast as
\begin{equation}
\label{eq:solVX}
\tilde v_x=-\frac iq\sum_{i=1}^3m^{N\Vert}_iC^{N\Vert}_ie^{m^{N\Vert}_iz},
\end{equation}
\begin{equation}
\label{eq:solVZ}
\tilde v_z=\sum_{i=1}^3C^{N\Vert}_ie^{m^{N\Vert}_iz},
\end{equation}
\begin{equation}
\label{eq:solP}
\tilde p=\sum_{i=1}^3A_iC^{N\Vert}_ie^{m^{N\Vert}_iz},
\end{equation}
\begin{equation}
\label{eq:solNX}
\tilde n_x=\sum_{i=1}^3B^\Vert_iC^{N\Vert}_ie^{m^{N\Vert}_iz},
\end{equation}
where
\begin{eqnarray}
\nonumber
\label{eq:A}
A_i=-\frac1{m^{N\Vert}_i}\left\{iq\frac{1+\lambda}2\right.\times
\qquad\qquad\qquad
\\
\nonumber
\times\left[K_1q^2-K_3\left(m^{N\Vert}_i\right)^2-\chi_a\mathcal H^2\right]
B^\Vert_i+
\\
+\left.
\left[i\omega\rho^N+\nu_3q^2
-\left(2\nu_1-\nu_3\right)\left(m^{N\Vert}_i\right)^2\right]
\right\},
\end{eqnarray}
\begin{equation}
\label{eq:BVert}
B^\Vert_i=\frac{i\gamma_1}{2q}\frac
{\left(1-\lambda\right)q^2-\left(1+\lambda\right)\left(m^{N\Vert}_i\right)^2}
{i\omega\gamma_1+K_1q^2-K_3\left(m^{N\Vert}_i\right)^2-\chi_a\mathcal H^2},
\end{equation}
\begin{equation}
\label{eq:mVert}
m^{N\Vert}_i=\left(\mu^\Vert_i\right)^{1/2},\quad\textrm{Re}\,m^{N\Vert}_i>0,
\end{equation}
and $\mu^\Vert_i$, $i=1,2,3$, are the roots of the cubic equation
\begin{equation}
\label{eq:Cubic}
a^\Vert_3\left(\mu^\Vert\right)^3+
a^\Vert_2\left(\mu^\Vert\right)^2+a^\Vert_1\mu^\Vert+a^\Vert_0=0,
\end{equation}
where
\begin{equation}
a^\Vert_3=\left(\frac{1-\lambda^2}4\gamma_1-\nu_3\right)K_3,
\end{equation}
\begin{eqnarray}
\nonumber
a^\Vert_2=i\omega\gamma_1\nu_3
-\left(\frac{1-\lambda^2}4\gamma_1-\nu_3\right)
\left(K_1q^2-\chi_a\mathcal H^2\right)
\\
+\left\{i\omega\rho^N
-\left[\frac{1+\lambda^2}2\gamma_1-2\left(\nu_1+\nu_2-\nu_3\right)\right]
q^2\right\}K_3,
\end{eqnarray}
\begin{eqnarray}
\nonumber
a^\Vert_1=-i\omega\gamma_1\left[i\omega\rho^N
+2\left(\nu_1+\nu_2-\nu_3\right)q^2
\right]
\\
\nonumber
-\left\{i\omega\rho^N
-\left[\frac{1+\lambda^2}2\gamma_1-2\left(\nu_1+\nu_2-\nu_3\right)\right]q^2
\right\}
\\
\nonumber
\times\left(K_1q^2-\chi_a\mathcal H^2\right)
\\
-\left[i\omega\rho^N
-\left(\frac{1-\lambda^2}4\gamma_1-\nu_3\right)q^2
\right]K_3q^2,
\end{eqnarray}
\begin{eqnarray}
\nonumber
a^\Vert_0&=&i\omega\gamma_1q^2\left(i\omega\rho^N+\nu_3q^2\right)
+\left[i\omega\rho^N\right.
\\
&-&\left.\left(\frac{1-\lambda^2}4\gamma_1-\nu_3\right)q^2
\right]\left(K_1q^2-\chi_a\mathcal H^2\right)q^2.
\end{eqnarray}

The quantities $C^{N\Vert}_i$ and $C^{N\bot}_i$ are functions of $q$ and
$\omega$ and are determined by the boundary conditions at the interface as
\begin{equation}
\label{eq:Cbot}
C^{N\bot}_1=
\frac{B^\bot_2\tilde v_y^S-\tilde n_y^S}{B^\bot_2-B^\bot_1},
\end{equation}
\begin{eqnarray}
\nonumber
C^{N\Vert}_1&=&\left[
iq\left(B^\Vert_3-B^\Vert_2\right)\tilde v_x^S
+\left(m^{N\Vert}_2-m^{N\Vert}_3\right)\tilde n_x^S
\right.
\\
\label{eq:CVert}
&+&\left.\left(m^{N\Vert}_3B^\Vert_2-m^{N\Vert}_2B^\Vert_3\right)\tilde v_z^S
\right]/\Delta,
\end{eqnarray}
where
\begin{eqnarray}
\nonumber
\Delta&=&\left(B^\Vert_3-B^\Vert_2\right)m^{N\Vert}_1
+\left(B^\Vert_1-B^\Vert_3\right)m^{N\Vert}_2
\\
\label{eq:Delta}
&+&\left(B^\Vert_2-B^\Vert_1\right)m^{N\Vert}_3,
\end{eqnarray}
and expressions for $C^{N\bot}_2$, $C^{N\Vert}_2$, and $C^{N\Vert}_3$ are
obtained from Eqs~(\ref{eq:Cbot}) and (\ref{eq:CVert}) by cyclic permutation
of subscript indices.


\section{\label{app:Dispersion}Explicit form of dispersion relation}

To write the explicit form of the dispersion relations (\ref{eq:DVert}) and
(\ref{eq:Dbot}), we recast equations (\ref{eq:Cbot}) and (\ref{eq:CVert}) in
form
\begin{equation}
C^{N\bot}_i=L^{\left(vy\right)}_i\tilde v_y+L^{\left(ny\right)}_i\tilde n_y,
\end{equation}
\begin{equation}
C^{N\Vert}_i=L^{\left(vx\right)}_i\tilde v_x+L^{\left(vz\right)}_i\tilde v_z,
+L^{\left(nx\right)}_i\tilde n_x
\end{equation}
where
\begin{eqnarray}
L^{\left(vy\right)}_1&=&\frac{B^\bot_2}{B^\bot_2-B^\bot_1},
\\
L^{\left(vy\right)}_2&=&\frac{B^\bot_1}{B^\bot_1-B^\bot_2},
\\
L^{\left(ny\right)}_1&=&\frac1{B^\bot_1-B^\bot_2},
\\
L^{\left(ny\right)}_2&=&\frac1{B^\bot_2-B^\bot_1},
\\
L^{\left(vx\right)}_1&=&\frac{iq\left(B^\Vert_3-B^\Vert_2\right)}\Delta,
\\
L^{\left(vx\right)}_2&=&\frac{iq\left(B^\Vert_1-B^\Vert_3\right)}\Delta,
\\
L^{\left(vx\right)}_3&=&\frac{iq\left(B^\Vert_2-B^\Vert_1\right)}\Delta,
\\
L^{\left(vz\right)}_1&=&
\frac{m^{N\Vert}_3B^\Vert_2-m^{N\Vert}_2B^\Vert_3}\Delta,
\\
L^{\left(vz\right)}_2&=&
\frac{m^{N\Vert}_1B^\Vert_3-m^{N\Vert}_3B^\Vert_1}\Delta,
\\
L^{\left(vz\right)}_3&=&
\frac{m^{N\Vert}_2B^\Vert_1-m^{N\Vert}_1B^\Vert_2}\Delta,
\\
L^{\left(nx\right)}_1&=&\frac{m^{N\Vert}_2-m^{N\Vert}_3}\Delta,
\\
L^{\left(nx\right)}_2&=&\frac{m^{N\Vert}_3-m^{N\Vert}_1}\Delta,
\\
L^{\left(nx\right)}_3&=&\frac{m^{N\Vert}_1-m^{N\Vert}_2}\Delta,
\end{eqnarray}
$\Delta$ is given by Eq.~(\ref{eq:Delta}), $B^\bot_i$ and $B^\Vert_i$ are given
by Eqs~(\ref{eq:Bbot}) and (\ref{eq:BVert}), $m^{N\bot}_i$ and $m^{N\Vert}_i$
are given by Eqs~(\ref{eq:mVert}) and (\ref{eq:mbot}), correspondingly.

Then the dispersion relation (\ref{eq:Dbot}) can be written as
\begin{equation}
\det M^\bot=0,
\end{equation}
where $M^\bot$ is $2\times2$ matrix of coefficients for equations
(\ref{eq:balanceVY}) and (\ref{eq:balanceNY})
\begin{equation}
M^\bot=
\begin{pmatrix}
M^\bot_{11} & M^\bot_{12}
\\
M^\bot_{21} & M^\bot_{22}
\end{pmatrix}
\end{equation}
with the following components:
\begin{eqnarray}
\nonumber
M^\bot_{11}=-\eta_sq^2-\eta m^I
+\sum_{i=1}^2\left\{
-\nu_3m^{N\bot}_i
+\frac{1+\lambda}2
\right.
\\
\times\left.\left[
K_3\left(m^{N\bot}_i\right)^2-K_2q^2+\chi_a\mathcal H^2
\right]B^\bot_i\right\}L^{\left(vy\right)}_i,
\end{eqnarray}
\begin{eqnarray}
\nonumber
M^\bot_{12}=\sum_{i=1}^2\left\{
-\nu_3m^{N\bot}_i+\frac{1+\lambda}2
\right.\times
\\
\times\left.\left[
K_3\left(m^{N\bot}_i\right)^2-K_2q^2+\chi_a\mathcal H^2
\right]B^\bot_i\right\}L^{\left(ny\right)}_i,
\end{eqnarray}
\begin{equation}
M^\bot_{21}=\sum_{i=1}^2
\left(K_3m^{N\bot}_i-W_0-i\omega\gamma^S_1\right)B^\bot_iL^{\left(vy\right)}_i,
\end{equation}
\begin{equation}
M^\bot_{22}=\sum_{i=1}^2
\left(K_3m^{N\bot}_i-W_0-i\omega\gamma^S_1\right)B^\bot_iL^{\left(ny\right)}_i.
\end{equation}

The dispersion relation (\ref{eq:DVert}) can be written as
\begin{equation}
\det M^\Vert=0,
\end{equation}
where $M^\Vert$ is $3\times3$ matrix of coefficients for equations
(\ref{eq:balanceVX}), (\ref{eq:balanceVZ}), and (\ref{eq:balanceNX})
\begin{equation}
M^\Vert=
\begin{pmatrix}
M^\Vert_{11} & M^\Vert_{12} & M^\Vert_{13}
\\
M^\Vert_{21} & M^\Vert_{22} & M^\Vert_{23}
\\
M^\Vert_{31} & M^\Vert_{32} & M^\Vert_{33}
\end{pmatrix}
\end{equation}
with the following components:
\begin{eqnarray}
\nonumber
M^\Vert_{11}=-\epsilon^\ast_0 q^2
-i\omega\eta\left(m^I+q\right)-
\\
\nonumber
-\frac\omega q\nu_3\sum_{i=1}^3\left[\left(m_i^{N\Vert}\right)^2+q^2\right]
L^{\left(vx\right)}_i
+i\omega\frac{1+\lambda}2\times
\\
\times\sum_{i=1}^3\left[K_3\left(m_i^{N\Vert}\right)^2-K_1q^2+\chi_a\mathcal H^2\right]B_i^\Vert
L^{\left(vx\right)}_i,
\end{eqnarray}
\begin{eqnarray}
\nonumber
M^\Vert_{12}=
-\omega\eta\left(m^I-q\right)-
\\
\nonumber
-\frac\omega q\nu_3\sum_{i=1}^3\left[\left(m_i^{N\Vert}\right)^2+q^2\right]
L^{\left(vz\right)}_i
+i\omega\frac{1+\lambda}2\times
\\
\times\sum_{i=1}^3\left[K_3\left(m_i^{N\Vert}\right)^2-K_1q^2+\chi_a\mathcal H^2\right]B_i^\Vert
L^{\left(vz\right)}_i,
\end{eqnarray}
\begin{eqnarray}
\nonumber
M^\Vert_{13}=
-\frac\omega q\nu_3\sum_{i=1}^3\left[\left(m_i^{N\Vert}\right)^2+q^2\right]
L^{\left(nx\right)}_i+
\\
\nonumber
{}+i\omega\frac{1+\lambda}2\times
\\
\times\sum_{i=1}^3\left[K_3\left(m_i^{N\Vert}\right)^2-K_1q^2+\chi_a\mathcal H^2\right]B_i^\Vert
L^{\left(nx\right)}_i,
\end{eqnarray}
\begin{eqnarray}
\nonumber
M^\Vert_{21}&=&
-iq^3\psi
-2\omega q\eta+\frac{i\omega^2\rho^I}{m^I-q}+
\\
&+&i\omega\sum_{i=1}^3\left(A_i-2\nu_1m_i^{N\Vert}\right)
L^{\left(vx\right)}_i,
\end{eqnarray}
\begin{eqnarray}
\nonumber
M^\Vert_{22}=
-g\left|\rho^N-\rho^I\right|
-\left(\sigma_0+W_0\right)q^2
-\kappa_0q^4+
\\
\label{eq:M22}
+\frac{\omega^2\rho^Im^I}{q\left(m^I-q\right)}
+i\omega\sum_{i=1}^3\left(A_i-2\nu_1m_i^{N\Vert}\right)
L^{\left(vz\right)}_i,
\end{eqnarray}
\begin{eqnarray}
\nonumber
M^\Vert_{23}&=&
\omega qW_0+
\\
&+&i\omega\sum_{i=1}^3\left(A_i-2\nu_1m_i^{N\Vert}\right)
L^{\left(nx\right)}_i,
\end{eqnarray}
\begin{eqnarray}
\nonumber
M^\Vert_{31}=
\sum_{i=1}^3\left\{\left(K_3m^{N\Vert}_i-W_0-i\omega\gamma^S_1\right)
B^\Vert_i\right.+
\\
+\frac q\omega\left.\left[
W_0+\kappa_0q^2-\psi m^{N\Vert}_i+i\omega\gamma^S_1
\right]\right\}L^{\left(vx\right)}_i,
\end{eqnarray}
\begin{eqnarray}
\nonumber
M^\Vert_{32}=
\sum_{i=1}^3\left\{\left(K_3m^{N\Vert}_i-W_0-i\omega\gamma^S_1\right)
B^\Vert_i\right.+
\\
+\frac q\omega\left.\left[
W_0+\kappa_0q^2-\psi m^{N\Vert}_i+i\omega\gamma^S_1
\right]\right\}L^{\left(vz\right)}_i,
\end{eqnarray}
\begin{eqnarray}
\nonumber
M^\Vert_{33}=
\sum_{i=1}^3\left\{\left(K_3m^{N\Vert}_i-W_0-i\omega\gamma^S_1\right)
B^\Vert_i\right.+
\\
+\frac q\omega\left.\left[
W_0+\kappa_0q^2-\psi m^{N\Vert}_i+i\omega\gamma^S_1
\right]\right\}L^{\left(nx\right)}_i.
\end{eqnarray}
Note that gravity $g$ has been incorporated into the dispersion relation by
adding the hydrostatic pressure term
$-g\left|\rho^N-\rho^I\right|$ to $M^\Vert_{22}$ (Eq.~(\ref{eq:M22})).

By setting $\nu_1=\nu_2=\nu_3$, and setting to zero quantities $K_1$, $K_2$,
$K_3$, $\lambda$, $\gamma_1$, $\gamma^S_{1\Vert}$, and $\chi_a$, specific to
nematic, and neglecting curvature contributions by setting to zero $\kappa_0$
and $\psi$, the dispersion relation is reduced to the well studied form for the
case of isotropic liquids
\cite{Kramer:1971-2097,Kats:1988-940,Earnshaw:1995-1087,Buzza:2002-8418}.



\end{document}